\documentclass[a4paper]{article}
\usepackage{a4wide}
\usepackage{float}
\usepackage[UKenglish]{babel}
\usepackage{authblk}
\usepackage{graphicx,subfigure}
\usepackage{epstopdf}
	\graphicspath{{./}{figures/}{figures/Aw_Rascle_conservativo/}{figures/Aw_Rascle_non_conservativo/}{figures/binary_control/}{figures/desired_speed_control/}}
\usepackage[colorlinks=true,linkcolor=blue,citecolor=red]{hyperref}
\usepackage{xcolor}
\usepackage{amsfonts,amsthm,mathtools,bbold}
\usepackage[inline]{enumitem}

\theoremstyle{remark}
\newtheorem{remark}{Remark}[section]

\newcommand{\A}{\mathbf{A}}
\newcommand{\abs}[1]{\left\lvert#1\right\rvert}
\DeclarePairedDelimiter\ave{\langle}{\rangle}
\newcommand{\E}{\mathbb{E}}
\newcommand{\Prob}{\operatorname{Prob}}
\newcommand{\R}{\mathbb{R}}
\newcommand{\U}{\mathbf{U}}
\newcommand{\cU}{\mathcal{U}}
\newcommand{\ucontr}{\mathfrak{u}}
\allowdisplaybreaks

\begin{document}

\title{Kinetic derivation of Aw-Rascle-Zhang-type traffic models with driver-assist vehicles}

\author[1]{Giacomo Dimarco}
\author[2]{Andrea Tosin}
\author[3]{Mattia Zanella}
\affil[1]{{\footnotesize Department of Mathematics and Computer Sciences, University of Ferrara, Italy}}
\affil[2]{{\footnotesize Department of Mathematical Sciences ``G. L. Lagrange'', Politecnico di Torino, Italy}}
\affil[3]{{\footnotesize Department of Mathematics ``F. Casorati'', University of Pavia, Italy}}
\date{}

\maketitle

\begin{abstract}
In this paper, we derive second order hydrodynamic traffic models from kinetic-controlled equations for driver-assist vehicles. At the vehicle level we take into account two main control strategies synthesising the action of adaptive cruise controls and cooperative adaptive cruise controls.  The resulting macroscopic dynamics fulfil the anisotropy condition introduced in the celebrated Aw-Rascle-Zhang model. Unlike other models based on heuristic arguments, our approach unveils the main physical aspects behind frequently used hydrodynamic traffic models and justifies the structure of the resulting macroscopic equations incorporating driver-assist vehicles. Numerical insights show that the presence of driver-assist vehicles produces an aggregate homogenisation of the mean flow speed, which may also be steered towards a suitable desired speed in such a way that optimal flows and traffic stabilisation are reached.

\medskip


\medskip

\end{abstract}

\section{Introduction}
In the field of vehicle automation, advanced driver assist technologies such as Adaptive Cruise Control (ACC) and Cooperative Adaptive Cruise Control (CACC) systems are likely to modify the classical paradigms of traffic dynamics to enhance driver safety. From the theoretical point of view, these technologies call for quantitative mathematical approaches which allow one to understand their aggregate effects and to design efficient next generation vehicles~\cite{Chow}. Among the most prominent goals that driver-assist technologies may pursue, we find both driver-oriented and flow-oriented issues. Driver-assist vehicles may indeed be programmed so as to either increase driver safety and comfort or optimise traffic governance tasks such as e.g., solving traffic congestion issues. For this, the development of mathematical tools able to model traffic flow incorporating vehicles equipped with driver-assistance systems is essential for an organic investigation and simulation of the potential of these technologies.

In the literature, these problems have been investigated under complementary mathematical and engineering perspectives. From the mathematical side, we mention for instance recent efforts based on the study of the controllability of partial differential equations modelling traffic flow dynamics. In~\cite{BKZ,chiarello2020MMS_preprint,piccoli2020ZAMP,tosin2018IFAC,tosin2019MMS} suitable control strategies have been introduced to mimic the local action of driver-assist or autonomous cars. In~\cite{tosin2020MCRF} robust control approaches aimed to enhance desired emerging features of automated traffic flows have been analysed. In~\cite{Garavello} a hybrid microscopic-macroscopic description is used to simulate a few individually controlled autonomous vehicles embedded in an aggregate flow of non-controlled vehicles modelled by the classical Lighthill-Whitham-Richards traffic equation~\cite{lighthill1955PRSL}. Optimal control methods have also been developed on networks~\cite{Goatin,Herty,Herty2}. Conversely, from the engineering point of view we recall some important contributions by Papageorgiou and co-authors. In~\cite{Pap1} microscopic vehicle-wise control models are reviewed while in~\cite{delis2015CMA} the contribution of adaptive cruise control systems is included in a second order hydrodynamic traffic model. The model is then extended in~\cite{Pap3} to the case of multilane traffic. These models take inspiration from the so-called gas-kinetic approach from~\cite{Helbing,Helbing2}. Extensions of controlled kinetic models to the multilane case has been studied in~\cite{BKZ,GPV}. Also field experiments have been developed recently to understand the aggregate effects of a small portion of automated vehicles in the traffic stream~\cite{Piccoli_exp}.

In this paper, we give a mathematical contribution within the conceptual framework of statistical mechanics, kinetic theory and multi-agent systems. Such a formalism allows us to bridge organically the microscopic scale of the vehicles, where driver-assist technologies act, and the macroscopic scale of the observable aggregate traffic phenomena. We mention that the control of multi-agent system has been recently explored as a natural follow-up of the description of their self-organisation abilities. Several methods have been developed for mean-field and kinetic equations~\cite{AHP,APZ14,Bensoussan,FPR} and for hyperbolic conservation laws~\cite{BandaHerty,Colombo,Cristiani_2015}. Here we present an approach based on a feedback formulation of the microscopic control, which may be effectively embedded into an Enskog-type kinetic description of traffic whence suitable hydrodynamic limits can then be computed.

In more detail, starting from a recent result~\cite{dimarco2020JSP} about the kinetic derivation of macroscopic traffic models of Aw-Rascle-Zhang (ARZ)-type~\cite{aw2000SIAP,zhang2002TRB}, we derive second order hydrodynamic models accounting for the presence of driver-assist vehicles. Second order macroscopic traffic models allow one to overcome several limitations of first order models and typically offer a richer set of more realistic flow dynamics~\cite{payne1971MMPS,piccoli2009ENCYCLOPEDIA}. Nevertheless, in the past they were the subject of a strong controversy as the first attempts of building second order hydrodynamic equations failed to reproduce the correct anisotropy of the interactions among the vehicles~\cite{Daganzo}. Indeed, in these models information can propagate both backwards and forwards, thereby leading to a situation in which vehicles ahead are influenced by those behind. This drawback was solved by Aw and Rascle in their seminal paper~\cite{aw2000SIAP} and independently Zhang~\cite{zhang2002TRB}. For this reason we will refer to this class of second order models as ARZ-type models. The ARZ correction guarantees that the movement of each vehicle affects only the vehicles behind but in the original work it is derived heuristically at the macroscopic scale, not from physical microscopic dynamics. Kinetic derivations of ARZ-type models have been proposed in the literature starting from the pioneering works~\cite{klar1997JSP,klar2000SIAP} up to the aforementioned recent one~\cite{dimarco2020JSP}. Taking advantage of such a kinetic derivation, here we define suitable control strategies at the vehicle level, that we subsequently upscale to the aggregate flow scale by means of suitable hydrodynamic limits to build macroscopic models. The key idea consists in considering both Boltzmann-type and Enskog-type kinetic descriptions leading to a complementary hydrodynamic scaling of two different components of the collision operator. The main contribution of the present work is therefore the derivation of microscopically controlled second order ARZ-type models from basic physical laws using a precise characterisation of the microscopic binary interactions among the vehicles and the mathematical tools of the kinetic theory. As a side result, we also identify the microscopic ingredients responsible for the macroscopic terms leading to the controlled dynamics.

To model the action of driver-assist vehicles we consider two alternative control strategies yielding two different sets of hydrodynamic equations. The first strategy is based on a binary control where we suppose that a driver-assist vehicle responds only locally, i.e. as a consequence of the behaviour of the vehicle ahead. This leads to a set of ARZ-type equations in which the pressure term responsible for the backward spreading of the information is consistently modified by the action of the control. The second strategy is based instead on a desired speed control in which a target speed is given as a function of the local level of traffic congestion. The resulting hydrodynamic model contains now a relaxation term in the momentum equation, which drives the mean flow towards the \textit{a priori} prescribed desired speed. It is worth to remarking that this control is consistent with existing macroscopic models where the action of controlled vehicles is heuristically modelled directly at the macroscopic scale~\cite{kolb2017NHM}, see also~\cite{greenberg2001SIAP} for an earlier approach. In both strategies the strength of the control depends on the \textit{penetration rate}, i.e. the fraction of vehicles equipped with a control device, a parameter that we explicitly obtain in our macroscopic equations from the upscaling of the microscopic vehicle dynamics.

The contents of the paper are specifically organised as follows. In Section~\ref{sect:Enskog}, we generalise the results obtained in~\cite{dimarco2020JSP} giving the necessary conditions needed to recover an ARZ-type model from an Enskog-type kinetic description. We stress, in particular, the role played by the random fluctuations in the driver behaviour. In Section~\ref{sect:bin_var}, we discuss the binary control strategy leading to an ARZ-type model with a modified pressure term with respect to the non-controlled case. In Section~\ref{sect:vd}, we derive instead the macroscopic model in the case of a desired speed control strategy depending on the local congestion of traffic. In Section~\ref{sect:numerics}, we present several numerical experiments. First, we discuss different discretisation techniques and we analyse the different results of the non-controlled system. Next, we investigate the trend of the system when the two control methods are active, showing that traffic is homogenised and the desired speed is reached. Finally, in Section~\ref{sect:conc} we present some concluding remarks and we briefly sketch further research perspectives.

\section{The Aw-Rascle-Zhang model from a Boltzmann-Enskog-type kinetic description}
\label{sect:Enskog}
We begin by introducing our kinetic description of traffic flow. Let $f(x,v,t)$ be the distribution function of vehicles located in $x\in\R$ and travelling at the (nondimensional) speed $v\in [0,\,1]$ at time $t\geq 0$. We assume that vehicles modify their speed via interactions with other vehicles located at a given constant headway $H>0$. At the kinetic level, these dynamics may be described by an Enskog-type equation, which in weak form reads
\begin{align}
	\begin{aligned}[b]
		\partial_t\int_0^1\varphi(v)f(x,v,t)\,dv &+ \partial_x\int_0^1v\varphi(v)f(x,v,t)\,dv \\
		&= \frac{1}{2}\int_0^1\int_0^1\ave{\varphi(v')-\varphi(v)}f(x,v,t)f(x+H,v_\ast,t)\,dv\,dv_\ast,
	\end{aligned}
	\label{eq:Enskog}
\end{align}
where $\varphi:[0,\,1]\to\R$ is a test function. Denoting by $v,\,v_\ast\in [0,\,1]$ the pre-interaction speeds of any two interacting vehicles, we prescribe the following \textit{follow-the-leader-inspired} binary laws ruling the speed changes after the interaction:
\begin{equation}
	\begin{cases}
		 v'=v+\gamma\lambda(\rho)(v_\ast-v)+D(v)\eta \\
		 v_\ast'=v_\ast,
	\end{cases}
	\label{eq:binary}
\end{equation}
where $v',\,v_\ast'$ are the post-interaction speeds and:
\begin{enumerate*}[label=(\roman*)]
\item $\gamma>0$ is a time-scale factor;
\item $\eta$ is a centred random variable with positive variance, i.e. $\ave{\eta}=0$ and $\ave{\eta^2}=\sigma^2>0$ with $\ave{\cdot}$ denoting expectation with respect to the law of $\eta$, which models random fluctuations in the driver behaviour;
\item $D:[0,\,1]\to\R_+$ is a function measuring the local relevance of the random fluctuations;
\item $\lambda(\rho)>0$ is, in the tradition of follow-the-leader traffic models~\cite{gazis1961OR}, the \textit{sensitivity} (or \textit{reactiveness}) of the drivers expressed as a function of the local vehicle density $\rho$ (see below for its precise definition).
\end{enumerate*}

The rules~\eqref{eq:binary} express the fact that the $v$-vehicle tends to adapt its speed to that of the $v_\ast$-vehicle ahead while the latter is not influenced by the $v$-vehicle behind. Suitable conditions should be placed on $D$, $\eta$ in order to guarantee that~\eqref{eq:binary} are physically admissible, in particular that $v',\,v_\ast'\in [0,\,1]$ for all $v,\,v_\ast\in [0,\,1]$. In~\cite{dimarco2020JSP,tosin2019MMS} the following sufficient conditions are provided:
$$	\begin{cases}
		\abs{\eta}\leq c(1-\gamma\lambda(\rho)) \\
		cD(v)\leq \min\{v,\,1-v\},
	\end{cases} $$
$c>0$ being an arbitrary constant, along with the requirement $\gamma\lambda(\rho)<1$ in order for the first condition to make sense. We will assume henceforth that this requirement is satisfied without further notice and that the non-negative function $D$ is not identically zero.

If, consistently with a hydrodynamic regime, we assume that the headway $H$ is small we can approximate
$$ f(x+H,v_\ast,t)\approx f(x,v_\ast,t)+H\partial_xf(x,v_\ast,t) $$
by a first order truncation of the Taylor expansion of $f$. Consequently, we also approximate~\eqref{eq:Enskog} as
\begin{align}
	\begin{aligned}[b]
		\partial_t\int_0^1\varphi(v)f(x,v,t)\,dv &+ \partial_x\int_0^1v\varphi(v)f(x,v,t)\,dv \\
		&= \frac{1}{2}\int_0^1\int_0^1\ave{\varphi(v')-\varphi(v)}f(x,v,t)f(x,v_\ast,t)\,dv\,dv_\ast \\
		&\phantom{=} +\frac{H}{2}\int_0^1\int_0^1\ave{\varphi(v')-\varphi(v)}f(x,v,t)\partial_xf(x,v_\ast,t)\,dv\,dv_\ast.
	\end{aligned}
	\label{eq:Enskog.approx}
\end{align}
Starting from~\eqref{eq:Enskog.approx}, hydrodynamic limits may be performed leading to second order macroscopic traffic models ruled by the microscopic dynamics~\eqref{eq:binary}. In the following, we summarise the formal procedure applied in~\cite{dimarco2020JSP}, which is based on the local equilibrium closure and will be the basis for the subsequent inclusion of driver-assist vehicles.

Let us introduce the following hyperbolic scaling of space and time:
\begin{equation}
	x\to\frac{2}{\epsilon}x, \qquad t\to\frac{2}{\epsilon}t,
	\label{eq:hydro_scaling}
\end{equation}
$0<\epsilon\ll 1$ being the analogous of the Knudsen number of the classical kinetic theory, i.e. a small parameter defining the hydrodynamic regime. Under such a scaling,~\eqref{eq:Enskog.approx} becomes
\begin{equation}
	\partial_t\int_0^1\varphi(v)f(x,v,t)\,dv+\partial_x\int_0^1v\varphi(v)f(x,v,t)\,dv
		=\frac{1}{\epsilon}(Q(f,f),\,\varphi)+\frac{H}{2}(Q(f,\partial_xf),\,\varphi),
	\label{eq:Enskog.eps}
\end{equation}
where $Q=Q(f,g)$ is the collision operator defined as
$$ (Q(f,g),\,\varphi):=\int_0^1\int_0^1\ave{\varphi(v')-\varphi(v)}f(x,v,t)g(x,v_\ast,t)\,dv\,dv_\ast $$
for every observable quantity $\varphi$. On the right-hand side of~\eqref{eq:Enskog.eps} we observe that, because of the presence of the space derivative of $f$, the time scale of the second collisional term is naturally different from that of the first collisional term. In particular, two time scales can be detected, which can be resolved by means of the following splitting, cf.~\cite{duering2007PHYSA}:
\begin{subequations}
\begin{align}
	&\partial_t\int_0^1\varphi(v)f(x,v,t)\,dv=\frac{1}{\epsilon}(Q(f,f),\,\varphi) \label{eq:split_1} \\
	&\partial_t\int_0^1\varphi(v)f(x,v,t)\,dv+\partial_x\int_0^1v\varphi(v)f(x,v,t)\,dv=\frac{H}{2}(Q(f,\partial_x f),\,\varphi). \label{eq:split_2}
\end{align}
Equation~\eqref{eq:split_1} describes now quick local interactions among the vehicles. In the hydrodynamic limit $\epsilon\to 0^+$, owing to the arbitrariness of $\varphi$, it produces
\begin{equation}
	Q(f,f)=0,
	\label{eq:split_3}
\end{equation}
\end{subequations}
whose solution yields the local equilibrium speed distribution, the so-called \textit{local Maxwellian} in the jargon of classical kinetic theory. Since, in view of~\eqref{eq:binary}, for every $\epsilon>0$ it results
$$ (Q(f,f),\,1)=(Q(f,f),\,v)=0, $$
the local density and the local mean speed of the vehicles, defined respectively as
$$ \rho(x,t):=\int_0^1f(x,v,t)\,dv, \qquad u(x,t):=\frac{1}{\rho(x,t)}\int_0^1vf(x,v,t)\,dv, $$
are conserved in time by the interactions. Consequently, the local Maxwellian resulting from~\eqref{eq:split_3} is spanned by $\rho$ and $u$. To stress this fact, we denote the local Maxwellian by $M_{\rho,u}=M_{\rho,u}(v)$. Specifically, we have that
$$ \int_0^1M_{\rho,u}(v)\,dv=\rho, \qquad \frac{1}{\rho}\int_0^1vM_{\rho,u}(v)\,dv=u. $$

Equation~\eqref{eq:split_2} expresses instead the slower transport of the local Maxwellian on the hydrodynamic spatio-temporal scale. This equation takes into account also the effect of the spatial dislocation of the interactions (right-hand side). When plugging $M_{\rho,u}$ into~\eqref{eq:split_2} with $(\varphi(v)=1,\,v)$ one is able to obtain the macroscopic spatio-temporal evolution of density $\rho$ and the mean speed $u$ of the vehicles. This gives
$$ (Q(M_{\rho,u},\partial_xM_{\rho,u}),\,1)=0, \qquad (Q(M_{\rho,u},\partial_xM_{\rho,u}),\,v)=\gamma\rho^2\lambda(\rho)\partial_xu, $$
and then
\begin{equation}
	\begin{cases}
		\partial_t\rho+\partial_x(\rho u)=0 \\
		\partial_t(\rho u)+\partial_x(\rho E)=\dfrac{\gamma H}{2}\rho^2\lambda(\rho)\partial_x u,
	\end{cases}
	\label{eq:macro_1}
\end{equation}
where we denote by
$$ E:=\frac{1}{\rho}\int_0^1v^2M_{\rho,u}(v)\,dv $$
the energy of the local equilibrium distribution. Recalling~\eqref{eq:split_3} for $f=M_{\rho,u}$ we deduce in particular $(Q(M_{\rho,u},M_{\rho,u}),\,v^2)=0$, whence the relationship among the energy of the local Maxwellian and the density and mean speed may be made explicit. This gives
\begin{equation}
	E=u^2+\frac{\sigma^2}{2\rho\gamma\lambda(\rho)(1-\gamma\lambda(\rho))}\int_0^1D^2(v)M_{\rho,u}(v)\,dv,
	\label{eq:E}
\end{equation}
whence we see that the equilibrium energy can be expressed as a function $E=E(\rho,u)$ of the hydrodynamic parameters $\rho$, $u$. In particular, $E\geq u^2$ consistently with the standard case of fluid dynamics where the total energy is the sum of the kinetic energy and the internal energy. Moreover, $E>u^2$ whenever $\sigma^2>0$.

Coming back to system~\eqref{eq:macro_1}, we observe that it may be fruitfully rewritten in quasilinear vector form as
$$ \partial_t\U+\A(\U)\partial_x\U=\mathbf{0}, $$
with $\U:=(\rho,\,u)^T$ and
$$ \A(\U):=
	\begin{pmatrix}
		u & \rho \\
		\partial_\rho E+\frac{T}{\rho} & \partial_uE-u-\frac{\gamma H}{2}\rho\lambda(\rho)
	\end{pmatrix}. $$
Here, $T:=E-u^2$ is the traffic temperature, viz. the variance of the vehicle speed at equilibrium. The eigenvalues of $\A(\U)$, representing the speeds of propagation of the small disturbances in the traffic flow, are given in this system by
$$ \mu_\pm:=\frac{1}{2}\partial_uE-\frac{\gamma H}{4}\rho\lambda(\rho)\pm\sqrt{\left(u-\frac{1}{2}\partial_uE+\frac{\gamma H}{4}\rho\lambda(\rho)\right)^2+T+\rho\partial_\rho E}. $$
We observe that if $T>0$ and $\partial_\rho E\geq 0$ then $\mu_\pm\in\R$, hence system~\eqref{eq:macro_1} is hyperbolic, and
$$ \mu_+>\frac{1}{2}\partial_uE-\frac{\gamma H}{4}\rho\lambda(\rho)+\abs{u-\frac{1}{2}\partial_uE+\frac{\gamma H}{4}\rho\lambda(\rho)}. $$
From here it follows that in the subregion of the state space $\{(\rho,\,u)\in\R_+\times [0,\,1]\}$ defined by the condition
\begin{equation}
	\partial_u E\leq 2u+\frac{\gamma H}{2}\rho\lambda(\rho)
	\label{eq:E_u}
\end{equation}
it results $\mu_+>u$. Therefore, in general, the second order hydrodynamic traffic model~\eqref{eq:macro_1} may violate the so-called \textit{Aw-Rascle (AR) condition}, which prescribes that the small disturbances of traffic should propagate at a speed at most equal to the mean speed of the flow and not faster, cf.~\cite{aw2000SIAP}. In the following remark we provide evidence of the fact that non-empty regions of the state space where $\mu_+>u$ might indeed exist.
\begin{remark}
From~\eqref{eq:E} we compute the derivatives of the energy $E$ with respect to the density $\rho$ and the mean speed $u$:
\begin{gather*}
	\partial_\rho E=\frac{\sigma^2}{2\rho\gamma\lambda(\rho)(1-\gamma\lambda(\rho))}\int_0^1D^2(v)\left[\partial_\rho M_{\rho,u}(v)
		-\left(\frac{\lambda'(\rho)(1-2\gamma\lambda(\rho))}{\lambda(\rho)(1-\gamma\lambda(\rho))}+\frac{1}{\rho}\right)M_{\rho,u}(v)\right]dv, \\[2mm]
	\partial_uE=2u+\frac{\sigma^2}{2\rho\gamma\lambda(\rho)(1-\gamma\lambda(\rho))}\int_0^1D^2(v)\partial_uM_{\rho,u}(v)\,dv.
\end{gather*}
Let us now assume that $\lambda$ does not depend on $\rho$. Then, the local Maxwellian takes the simpler form $M_{\rho,u}(v)=\rho g_u(v)$, where $g_u$ is a probability density independent of $\rho$ with mean $u$. The reason is that, in this particular case, the interaction rules~\eqref{eq:binary} do not depend on $\rho$ any more, hence so does also the probability distribution of $v$ (in other words, different densities produce simply self-similar local Maxwellians). Consequently, from the formulas above we obtain that $\partial_\rho E=0$. Moreover, equation~\eqref{eq:E_u} becomes
$$ \int_0^1D^2(v)\partial_ug_u(v)\,dv\leq\rho^2\frac{\gamma^2\lambda^2(1-\gamma\lambda)H}{\sigma^2}. $$
Since the left-hand side is independent of $\rho$ while the right-hand side is proportional to $\rho^2$, this condition, hence also $\mu_+>u$, may well be satisfied in a non-empty subregion of the state space where $\rho$ is large enough.
\end{remark}

From these results it is clear that if $T>0$ the AR condition may be violated in general. If conversely $T=0$ then $E=u^2$, therefore $\partial_\rho E=0$ and
$$ \mu_-=u-\frac{\gamma H}{2}\rho\lambda(\rho), \qquad \mu_+=u $$
fulfils the AR condition for all $(\rho,\,u)\in\R_+\times [0,\,1]$. The hydrodynamic model resulting in this case from~\eqref{eq:macro_1} is an ARZ model~\cite{aw2000SIAP,zhang2002TRB}, which is more often written in the form
\begin{equation}
	\begin{cases}
		\partial_t\rho+\partial_x(\rho u)=0 \\
		\partial_t(u+p(\rho))+u\partial_x(u+p(\rho))=0
	\end{cases}
	\label{eq:AR}
\end{equation}
with the \textit{traffic pressure} $p=p(\rho)$ defined by the relationship
\begin{equation}
	p'(\rho):=\frac{\gamma H}{2}\lambda(\rho).
	\label{eq:AR.p}
\end{equation}

The thermodynamic assumption $T=0$ underlying the derivation of~\eqref{eq:AR} is equivalent to the microscopic assumption $\sigma^2=0$, cf.~\eqref{eq:E}. In other words, the ARZ model is obtained as the hydrodynamic limit of the interacting particle model~\eqref{eq:binary} only if the latter is deterministic, i.e. if randomness in the driver behaviour is disregarded ($\eta\equiv 0$). This implies also that the local Maxwellian is the monokinetic one $M_{\rho,u}(v)=\rho\delta(v-u)$, where $\delta(v-u)$ denotes the Dirac delta distribution centred in $v=u$. Finally, it is worth pointing out that the Boltzmann-Enskog kinetic description is essential to recover the ARZ model. Indeed, in the case $H=0$ equation~\eqref{eq:Enskog} reduces to a Boltzmann-type equation. Then, from~\eqref{eq:macro_1} one obtains either a pressureless hydrodynamic model featuring two coincident eigenvalues $\mu_\pm=u$ if $T=0$ or a macroscopic model violating the AR condition in the subregion of the state space where $\partial_uE\leq 2u$ if $T>0$, $\partial_\rho E\geq 0$.

\begin{remark}
The results presented in this section generalise those presented in~\cite{dimarco2020JSP}, which were obtained in the particular case of a family of beta-type local Maxwellians stemming from~\eqref{eq:binary}-\eqref{eq:split_1} in the \textit{quasi-invariant interaction regime}. The concept of quasi-invariant interactions was introduced in the kinetic theory of multi-agent systems in~\cite{toscani2006CMS}, taking inspiration from the \textit{grazing collisions} of classical kinetic theory~\cite{villani1998PhD,villani1998ARMA}. We refer the interested reader to these contributions for details.
\end{remark}

Since we are interested in second order hydrodynamic models fulfilling the AR condition, we will henceforth fix $\eta\equiv 0$ in the interaction rules. 

\section{Binary variance control and modified Aw-Rascle-Zhang model}
\label{sect:bin_var}
We discuss in this section a first control strategy, which may be used to model ACC devices. A driver-assist vehicle responds locally to the actions of its driver to optimise the driving style, taking into account information coming from the vehicles ahead. Such an optimisation typically aims to mitigate driving risks, for instance by keeping a safety distance or reducing the speed gap from the leading vehicle. A consistent way to describe a driver-assist vehicle of this kind is therefore through a control of the basic binary dynamics~\eqref{eq:binary}:
\begin{equation}
	\begin{cases}
		v'=v+\gamma\bigl(\lambda(\rho)(v_\ast-v)+\Theta\ucontr\bigr) \\
		v_\ast'=v_\ast,
	\end{cases}
	\label{eq:binary.u}
\end{equation}
where, owing to the results of Section~\ref{sect:Enskog}, we have set $\eta\equiv 0$. Here, $\ucontr$ is the control of the interaction operated by the driver-assist device, which, in view of the discussion above, we imagine in \textit{feedback form}, i.e. $\ucontr=\ucontr(v,v_\ast)$. Furthermore, $\Theta\sim\operatorname{Bernoulli}(q)$ is a random variable discriminating whether a randomly selected vehicle in the traffic flow is ($\Theta=1$) or is not ($\Theta=0$) equipped with driver-assist technology. The parameter
$$ q:=\Prob(\Theta=1)\in [0,\,1] $$
gives then the fraction of driver-assist vehicles in the traffic stream. Commonly known as the \textit{penetration rate} in the transportation engineering literature, $q$ is nowadays estimated within a benchmark range of $5$-$10\%$~\cite{schoettle2015TR}.

The control $\ucontr$ is chosen as the minimiser of a prescribed \textit{instantaneous} cost functional $J=J(v',v_\ast',\ucontr)$. Therefore, the \textit{instantaneous} optimal control $\ucontr^\ast$ is such that
$$ \ucontr^\ast=\operatorname*{arg\,min}_{\ucontr\in\cU}J(v',v_\ast',\ucontr) $$
subject to~\eqref{eq:binary.u}, $\cU$ being the set of admissible controls. In our context, the admissibility of a control $\ucontr$ is essentially related to the physical admissibility of the resulting binary rules~\eqref{eq:binary.u}, i.e. $\cU=\{\ucontr\,:\,v'\in [0,\,1]\}$.

As far as the choice of the functional $J$ is concerned, similarly to~\cite{tosin2018IFAC,tosin2019MMS} we are interested in steering the post-interaction speed $v'$ towards a prescribed target speed. A concrete possibility is to take such a target speed coinciding with $v_\ast'$, which corresponds to aligning $v'$ to the speed of the leading vehicle thereby aiming at reducing the local speed fluctuations produced by the interactions. Therefore we consider
\begin{equation}
	J(v',v_\ast',\ucontr)=\frac{1}{2}\left[(v_\ast'-v')^2+\nu\ucontr^2\right],
	\label{eq:J.v*}
\end{equation}
where the first term penalises too different post-interaction speeds and may be conceptually assimilated to the \textit{binary variance} of the speeds of the interacting vehicles. The second term penalises instead too strong controls. In particular, the parameter $\nu>0$ may be understood as the cost of the control. We may determine the optimal control through a standard approach based on Lagrange multipliers. We define the Lagrangian
$$ \mathcal{L}(v^\prime,\ucontr,\beta)=J(v^\prime,v_\ast,\ucontr)+\beta[v^\prime-v-\gamma(\lambda(\rho)(v_\ast-v)+\Theta\ucontr)], $$
where $\beta\in\R$ is the Lagrange multiplier associated with the first constraint in~\eqref{eq:binary.u}. Notice that the second constraint has been directly imposed in the expression of the functional $J$. The optimality conditions are then given by
$$	\begin{cases}
		\dfrac{\partial\mathcal{L}}{\partial v^\prime}=-(v_\ast-v')+\beta=0 \\[0.25cm]
		\dfrac{\partial\mathcal{L}}{\partial\ucontr}=\nu\ucontr-\beta\gamma\Theta=0 \\[0.25cm]
		\dfrac{\partial\mathcal{L}}{\partial\beta}=v^\prime-v-\gamma(\lambda(\rho)(v_\ast-v)+\Theta\ucontr)=0,
	\end{cases}
$$
whence we get the optimal binary control in feedback form 
$$ \ucontr^\ast=\frac{\gamma\Theta(1-\gamma\lambda(\rho))}{\nu+\gamma^2\Theta^2}(v_\ast-v) $$
which may be directly substituted in~\eqref{eq:binary.u} to give the following instantaneously controlled binary interactions:
\begin{equation}
	\begin{cases}
		v'=v+\gamma\dfrac{\nu\lambda(\rho)+\gamma\Theta^2}{\nu+\gamma^2\Theta^2}(v_\ast-v) \\
		v_\ast'=v_\ast.
	\end{cases}
	\label{eq:binary.u*}
\end{equation}
It can be checked that the condition $\gamma\lambda(\rho)<1$ guarantees the physical admissibility also of these new interaction rules, hence in particular that $\ucontr^\ast\in\cU$.

The Boltzmann-Enskog kinetic description of the particle system ruled by~\eqref{eq:binary.u*} is provided, under the hydrodynamic scaling~\eqref{eq:hydro_scaling} of space and time and within the approximation of $H$ small, by~\eqref{eq:Enskog.eps} with a suitable modification of the collision operator $Q$, which now averages the microscopic interactions also with respect to the randomness induced by $\Theta$:
$$ (Q(f,g),\,\varphi):=\E_\Theta\!\left[\int_0^1\int_0^1(\varphi(v')-\varphi(v))f(x,v,t)g(x,v_\ast,t)\,dv\,dv_\ast\right], $$
where $\E_\Theta[\cdot]$ denotes the expectation with respect to the law of $\Theta$. At the same time, we notice that the $\eta$-average $\ave{\cdot}$ is no longer needed, being the stochastic fluctuations of the driver behaviour set to zero.

For $\epsilon>0$ small we can perform the same splitting~\eqref{eq:split_1}-\eqref{eq:split_2} and subsequently the same formal limit procedure. In particular, from~\eqref{eq:split_1} we observe that $\varphi(v)=1,\,v$ are still collisional invariants because $(Q(f,f),\,1)=(Q(f,f),\,v)=0$ for all $\epsilon>0$. This indicates that in the hydrodynamic limit $\epsilon\to 0^+$ the local Maxwellian solving~\eqref{eq:split_3} is still parametrised by $\rho$, $u$, i.e. $M=M_{\rho,u}(v)$.
The energy of the local Maxwellian can be found from
$$ 0=(Q(M_{\rho,u},M_{\rho,u}),\,v^2)=2\gamma\rho^2(1-\gamma\lambda(\rho))\left(q\frac{\nu^2\lambda(\rho)+\gamma\nu}{(\nu+\gamma^2)^2}+(1-q)\lambda(\rho)\right)(u^2-E) $$
whence $E=u^2$, i.e. $T=0$, which yields finally $M_{\rho,u}(v)=\rho\delta(v-u)$.

Plugging this result in~\eqref{eq:split_2} to obtain the macroscopic transport of the hydrodynamic parameters $\rho$, $u$ we have
\begin{equation}
	\partial_t(\rho\varphi(u))+\partial_x(\rho u\varphi(u))=\frac{H}{2}(Q(M_{\rho,u},\partial_xM_{\rho,u}),\,\varphi),
	\label{eq:ARZ.controlled-weak}
\end{equation}
with in particular $(Q(M_{\rho,u},\partial_xM_{\rho,u}),\,1)=0$ and
$$ (Q(M_{\rho,u},\partial_xM_{\rho,u}),\,v)=\gamma\left(1+q\gamma\frac{1-\gamma\lambda(\rho)}{(\nu+\gamma^2)\lambda(\rho)}\right)\rho^2\lambda(\rho)\partial_xu. $$

On the whole, from~\eqref{eq:ARZ.controlled-weak} with $\varphi(v)=1,\,v$ we recover the second order hydrodynamic model
\begin{equation}
	\begin{cases}
		\partial_t\rho+\partial_x(\rho u)=0 \\
		\partial_t(u+P(\rho))+u\partial_x(u+P(\rho))=0
	\end{cases}
	\label{eq:ARZ.controlled.bin_var}
\end{equation}
with
$$ P'(\rho):=\frac{\gamma H}{2}\left(1+q\gamma\frac{1-\gamma\lambda(\rho)}{(\nu+\gamma^2)\lambda(\rho)}\right)\lambda(\rho). $$
Since $\gamma\lambda(\rho)<1$ we have $P'(\rho)>p'(\rho)$, where $p(\rho)$ is the traffic pressure of the original (uncontrolled) ARZ model. In essence, with the binary variance control we recover the ARZ model with a higher traffic pressure, the increase in the pressure being proportional to the penetration rate $q$. From the results of Section~\ref{sect:Enskog} we deduce straightforwardly that model~\eqref{eq:ARZ.controlled.bin_var} complies with the AR condition.

Notice that for either $q=0$, i.e. if there are no driver-assist vehicles in the traffic stream, or $\nu\to+\infty$, i.e. if the cost of the control is so large that the control cannot be implemented in practice, we consistently recover the uncontrolled ARZ model.

\section{Desired speed control and relaxation hydrodynamic equations}
\label{sect:vd}
We consider now a different control strategy, which targets a \textit{desired speed} $v_d(\rho)\in [0,\,1]$ based on the local traffic congestion. This control strategy shares some similarities with driver-assist technologies based on a \textit{non-local} communication among vehicles in the traffic stream. Indeed, at the microscopic level of single vehicles the evaluation of the traffic density requires non-local information. The function $v_d(\rho)$ may be further chosen so as to optimise macroscopic traffic properties as recently proposed in~\cite{chiarello2020MMS_preprint}. The main difference with the binary variance control strategy of Section~\ref{sect:bin_var} is that in this case the target speed is not linked to microscopic vehicle dynamics but to aggregate traffic properties. Indeed $v_d(\rho)$ may be seen, on the whole, as an external input compared to the characteristic scale at which driver-assist vehicles operate. For this reason, it is convenient to distinguish two types of speed updates experienced parallelly by a vehicle:
\begin{subequations}
\begin{enumerate}[label=\roman*)]
\item one due to regular interactions with other vehicles, which follows the basic (uncontrolled) binary rules
\begin{equation}
	\begin{cases}
		 v'=v+\gamma\lambda(\rho)(v_\ast-v) \\
		 v_\ast'=v_\ast;
	\end{cases}
	\label{eq:binary.vd.1}
\end{equation}
\item another one, which does not consist in an interaction with another vehicle but in the speed modification by the driver-assist control $\ucontr$ based on the knowledge of external information:
\begin{equation}
	v''=v+\gamma\Theta\ucontr.
	\label{eq:binary.vd.2}
\end{equation}
In particular, we expect $\ucontr=\ucontr(v,\rho)$.
\end{enumerate}
The instantaneous optimal control is then determined from the minimisation of a cost functional similar to~\eqref{eq:J.v*} with $v_\ast'$ replaced by $v_d(\rho)$:
$$ J(v'',\ucontr)=\frac{1}{2}\left[(v_d(\rho)-v'')^2+\nu\ucontr^2\right] $$
constrained to~\eqref{eq:binary.vd.2}. Via a Lagrange multiplier approach analogous to the one followed in Section~\ref{sect:bin_var} we obtain the feedback optimal control
$$ \ucontr^\ast=\frac{\gamma\Theta}{\nu+\gamma^2\Theta^2}(v_d(\rho)-v), $$
which specialises~\eqref{eq:binary.vd.2} into
\begin{equation}
	v''=v+\frac{\gamma^2\Theta^2}{\nu+\gamma^2\Theta^2}(v_d(\rho)-v).
	\label{eq:binary.vd.3}
\end{equation}
It can be readily checked that $v''\in [0,\,1]$ for all $v\in [0,\,1]$, hence that $\ucontr^\ast$ is admissible, under the natural assumptions $v_d(\rho)\in [0,\,1]$ and $\gamma,\,\nu,\,\Theta\geq 0$. No further constraints on the parameters are needed in this case. We also observe that if the cost of the control is negligible, i.e. $\nu\to 0^+$, the updated speed relaxes immediately towards the desired speed $v_d(\rho)$.
\end{subequations}

The Boltzmann-Enskog kinetic equation corresponding to the particle dynamics~\eqref{eq:binary.vd.1}-\eqref{eq:binary.vd.3} includes now two integral terms on the right-hand side:
\begin{align}
	\begin{aligned}[b]
		\partial_t\int_0^1\varphi(v)f(x,v,t)\,dv &+ \partial_x\int_0^1v\varphi(v)f(x,v,t)\,dv \\
		&= \frac{1}{4}\int_0^1\int_0^1(\varphi(v')-\varphi(v))f(x,v,t)f(x+H,v_\ast,t)\,dv\,dv_\ast \\
		&\phantom{=} +\frac{\epsilon}{2}\E_\Theta\!\left[\int_0^1(\varphi(v'')-\varphi(v))f(x,v,t)\,dv\right].
	\end{aligned}
	\label{eq:Enskog.vd}
\end{align}
The first one, which reproduces the collision operator of Section~\ref{sect:Enskog} (without $\eta$), accounts for the average effect of the interactions~\eqref{eq:binary.vd.1}. The second one describes instead the simultaneous average contribution of process~\eqref{eq:binary.vd.3} to the speed variations. The parameter $\epsilon>0$ is the frequency of process~\eqref{eq:binary.vd.3}, which we assume to be much lower than that of process~\eqref{eq:binary.vd.1} following the idea that vehicle-to-vehicle interactions happen more frequently than speed changes dictated by the alignment to an external recommended speed. Therefore we consider $\epsilon\ll 1$.

Performing the approximation~\eqref{eq:Enskog.approx} of the collision operator and using $\frac{\epsilon}{2}$ as the Knudsen number in the hydrodynamic scaling~\eqref{eq:hydro_scaling} of space and time, we split~\eqref{eq:Enskog.vd} as
\begin{subequations}
\begin{align}
	&\partial_t\int_0^1\varphi(v)f(x,v,t)\,dv=\frac{1}{\epsilon}(Q(f,f),\,\varphi) \label{eq:split.vd_1} \\
	&\partial_t\int_0^1\varphi(v)f(x,v,t)\,dv+\partial_x\int_0^1v\varphi(v)f(x,v,t)\,dv
		=\frac{H}{4}(Q(f,\partial_x f),\,\varphi)+(R(f),\,\varphi), \label{eq:split.vd_2}
\end{align}
\end{subequations}
where the operators $Q$, $R$ are defined by
\begin{align*}
	& (Q(f,g),\,\varphi):=\int_0^1\int_0^1(\varphi(v')-\varphi(v))f(x,v,t)g(x,v_\ast,t)\,dv\,dv_\ast \\
	& (R(f),\,\varphi):=2\E_\Theta\!\left[\int_0^1(\varphi(v'')-\varphi(v))f(x,v,t)\,dv\right]
\end{align*}
for every observable quantity $\varphi$.

Equation~\eqref{eq:split.vd_1} with the binary interactions~\eqref{eq:binary.vd.1} is essentially the same as~\eqref{eq:split_1} with the binary interactions~\eqref{eq:binary} (and $\eta\equiv 0$). Therefore from Section~\ref{sect:Enskog} we know that the local Maxwellian resulting from the collision step~\eqref{eq:split.vd_1} is $M_{\rho,u}(v)=\rho\delta(v-u)$. Next, considering that
\begin{align*}
	& (Q(M_{\rho,u},\partial_xM_{\rho,u}),\,\varphi)=\gamma\rho^2\lambda(\rho)\partial_x\varphi(u), \\[1mm]
	& (R(M_{\rho,u}),\,\varphi)=2q\rho\!\left[\varphi\!\left(u+\frac{\gamma^2}{\nu+\gamma^2}(v_d(\rho)-u)\right)-\varphi(u)\right]
\end{align*}
and taking $\varphi(v)=1,\,v$ together with $f=M_{\rho,u}$ in~\eqref{eq:split.vd_2} we obtain the hydrodynamic system
\begin{equation}
	\begin{cases}
		\partial_t\rho+\partial_x(\rho u)=0 \\
		\partial_t(u+p(\rho))+u\partial_x(u+p(\rho))=\dfrac{2q\gamma^2}{\nu+\gamma^2}(v_d(\rho)-u),
	\end{cases}
	\label{eq:Greenberg}
\end{equation}
where $p(\rho)$ is half the traffic pressure of the ARZ model defined by~\eqref{eq:AR.p}. Hence~\eqref{eq:Greenberg} is an ARZ-type model \textit{with relaxation}, the relaxation term on the right-hand side of the second equation expressing the aggregate effect of driver-assist vehicles implementing a desired speed control strategy. The coefficient
\begin{equation}
	\tau:=\frac{\nu+\gamma^2}{2q\gamma^2}
	\label{eq:tau}
\end{equation}
is the \textit{relaxation time}, namely the characteristic time needed for the mean speed $u$ to relax locally towards the recommended speed $v_d(\rho)$.

\begin{remark}
From~\eqref{eq:tau} we observe that $\tau$ diminishes, i.e. the relaxation is quicker, if either $q$ is large, i.e. there is a high percentage of driver-assist vehicles in the traffic stream, or $\nu$ is small, i.e. the control is cheap (viz. stronger). Nevertheless we also observe that $\tau$ is bounded away from zero, indeed $\tau\geq\frac{1}{2}$ for all $\gamma,\,\nu,\,q\geq 0$. The minimum value $\tau=\frac{1}{2}$ is attained for $\nu=0$ and $q=1$ with $\gamma>0$. This means that, unlike the microscopic dynamics~\eqref{eq:binary.vd.3}, even a zero-cost control implemented on all vehicles cannot produce macroscopically an instantaneous local relaxation of $u$ towards $v_d(\rho)$, which would require instead $\tau\to 0^+$ in~\eqref{eq:Greenberg}.

The mathematical reason behind this difference between the microscopic and the macroscopic relaxation processes relies on the assumption, in the kinetic equation~\eqref{eq:Enskog.vd}, that the driver-assist control acts on a much slower time scale than the vehicle interactions. This relegates the relaxation operator $R$ to the splitting step~\eqref{eq:split.vd_2}, i.e. to the hydrodynamic scale, with no possibility to affect the local Maxwellian resulting from the splitting step~\eqref{eq:split.vd_1}.

However, if in~\eqref{eq:Enskog.vd} one drops the coefficient $\epsilon$ on the right-hand side, thereby assuming that the driver-assist control acts on the same time scale as the vehicle interactions, then the operator $R$ enters the splitting step~\eqref{eq:split.vd_1} and produces $u=v_d(\rho)$ at the local equilibrium. In particular, it may be checked that in that step $u$ converges exponentially fast in time to $v_d(\rho)$ with characteristic time $\tau$. Consequently the local Maxwellian becomes $M_{\rho}(v)=\rho\delta(v-v_d(\rho))$, to which there corresponds a \textit{first order} hydrodynamic model ruled by the Lighthill-Whitham-Richards-type equation~\cite{lighthill1955PRSL,richards1956OR}
$$ \partial_t\rho+\partial_x(\rho v_d(\rho))=0. $$
Such a hydrodynamic model is justified if the local relaxation of $u$ towards $v_d(\rho)$ in~\eqref{eq:split.vd_1} is so quick, viz. $\tau$ is so small, that it may be considered instantaneous on the macroscopic scale. Nevertheless, since as observed before $\tau$ is bounded away from zero, a second order hydrodynamic model with finite (i.e. non-infinitesimal) relaxation time of $u$ towards $v_d(\rho)$ seems to be more consistent from both the mathematical and the technological points of view. This further supports \textit{a posteriori} the choice to introduce the coefficient $\epsilon$ on the right-hand side of the kinetic equation~\eqref{eq:Enskog.vd}.
\label{rem:relaxation}
\end{remark}

From the results of Section~\ref{sect:Enskog} we deduce immediately that model~\eqref{eq:Greenberg} is hyperbolic and fulfils the AR condition in the whole state space $\{(\rho,\,u)\in\R_+\times [0,\,1]\}$.
A model similar to~\eqref{eq:Greenberg} was proposed in~\cite{greenberg2001SIAP,rascle2002MCM} as a heuristic extension of the ARZ model to incorporate the tendency of vehicles to travel locally at the maximum possible speed. The latter was in turn understood as a function of the local traffic density. The same model~\cite{greenberg2001SIAP,rascle2002MCM} was also applied in~\cite{kolb2017NHM} to traffic problems on road networks with ramps. Interestingly, here we have recovered a class of macroscopic models with the same features as~\cite{greenberg2001SIAP,rascle2002MCM} as the result of a consistent derivation from first principles. Our derivation highlights in particular the fundamental microscopic processes at the basis of this hydrodynamic model and their unobvious relationships (see Remarks~\ref{rem:relaxation} above and~\ref{rem:I-II_order} below). Moreover, it justifies the use of relaxation terms in second order macroscopic traffic models to reproduce \textit{certain} actions of driver-assist vehicles. In this respect, it is worth mentioning that such terms have been heuristically included in other second order macroscopic traffic models, which however do not fulfil the AR condition, see e.g.~\cite{delis2015CMA}.

\begin{remark}
Implementing the desired speed control into binary interactions of the form~\eqref{eq:binary.u} would yield in turn a first order hydrodynamic traffic model for physically different but mathematically analogous reasons to those discussed in Remark~\ref{rem:relaxation}. The crucial point is that also in this case the microscopic dynamics entering the splitting step~\eqref{eq:split.vd_1} would \textit{not} conserve the local mean speed, which would be then locally a function of the traffic density at equilibrium, cf.~\cite{piccoli2020ZAMP,tosin2019MMS}. The different structure of rules~\eqref{eq:binary.vd.1}-\eqref{eq:binary.vd.2}, along with the assumption that the two processes take place at different rates, are the microscopic origin of the conservation of the mean speed in the local interactions, which finally produces a second order hydrodynamic model.
\label{rem:I-II_order}
\end{remark}

\subsection{Mixed control strategies}
As a by-product of the results of Sections~\ref{sect:bin_var},~\ref{sect:vd}, we can finally and easily address the case of a traffic stream incorporating both vehicles implementing a binary variance control strategy and vehicles implementing a desired speed control strategy. To this purpose, we consider the following interaction rules:
\begin{subequations}
\begin{gather}
	\begin{cases}
		 v'=v+\gamma\bigl(\lambda(\rho)(v_\ast-v)+\Theta_1\ucontr_1\bigr) \\
		 v_\ast'=v_\ast,
	\end{cases}
	\label{eq:mixed.u-1}
	\\
	v''=v+\gamma\Theta_2\ucontr_2,
	\label{eq:mixed.u-2}
\end{gather}
\end{subequations}
where $\Theta_1\sim\operatorname{Bernoulli}(q_1)$ and $\Theta_2\sim\operatorname{Bernoulli}(q_2)$ discriminate whether a vehicle is or is not equipped with a binary variance control device and a desired speed control device, respectively. The parameters $q_1,\,q_2\in [0,\,1]$ are the corresponding probabilities. Assuming $\Theta_1$, $\Theta_2$ independent, the probability that a vehicle is equipped with both control systems is $q_1q_2$ while the probability that a vehicle is equipped with only one control system is $q_1(1-q_2)+(1-q_1)q_2$. Finally, the probability that a vehicle is not equipped with any driver-assist system is $(1-q_1)(1-q_2)$. Clearly, with $q_1=0$ or $q_2=0$ we recover the cases discussed in the previous sections.

Using the same cost functionals as in Sections~\ref{sect:bin_var},~\ref{sect:vd}, the optimal feedback controls $\ucontr^\ast_1$, $\ucontr^\ast_2$ are given by
$$ \ucontr^\ast_1=\frac{\gamma\Theta_1(1-\gamma\lambda(\rho))}{\nu_1+\gamma^2\Theta_1^2}(v_\ast-v),
	\qquad \ucontr^\ast_2=\frac{\gamma\Theta_2}{\nu_2+\gamma^2\Theta_2^2}(v_d(\rho)-v), $$
$\nu_1,\,\nu_2>0$ being the respective control penalizations.

The Boltzmann-Enskog kinetic equation corresponding to the particle dynamics~\eqref{eq:mixed.u-1}-\eqref{eq:mixed.u-2} is now
\begin{align*}
	\partial_t\int_0^1\varphi(v)f(x,v,t)\,dv &+ \partial_x\int_0^1v\varphi(v)f(x,v,t)\,dv \\
	&= \frac{1}{4}\E_{\Theta_1}\!\left[\int_0^1\int_0^1(\varphi(v')-\varphi(v))f(x,v,t)f(x+H,v_\ast,t)\,dv\,dv_\ast\right] \\
	&\phantom{=} +\frac{\epsilon}{2}\E_{\Theta_2}\!\left[\int_0^1(\varphi(v'')-\varphi(v))f(x,v,t)\,dv\right],
\end{align*}
still assuming that~\eqref{eq:mixed.u-2} takes place at a smaller rate $0<\epsilon\ll 1$ than~\eqref{eq:mixed.u-1}. From this, by the hydrodynamic scaling of space and time and the subsequent splitting in the hydrodynamic limit $\epsilon\to 0^+$ we recover the macroscopic ARZ-type model
\begin{equation}
	\begin{cases}
		\partial_t\rho+\partial_x(\rho u)=0 \\
		\partial_t(u+P(\rho))+u\partial_x(u+P(\rho))=\dfrac{2q_2\gamma^2}{\nu_2+\gamma^2}(v_d(\rho)-u)
	\end{cases}
	\label{eq:Greenberg+bin_var}
\end{equation}
with
$$ P'(\rho):=\frac{\gamma H}{4}\left(1+q_1\gamma\frac{1-\gamma\lambda(\rho)}{(\nu_1+\gamma^2)\lambda(\rho)}\right)\lambda(\rho). $$

Writing the second equation of system~\eqref{eq:Greenberg+bin_var} as
$$ \partial_t(\rho u)+\partial_x(\rho u^2)=\frac{\gamma H}{4}\left(1+q_1\gamma\frac{1-\gamma\lambda(\rho)}{(\nu_1+\gamma^2)\lambda(\rho)}\right)\rho^2\lambda(\rho)\partial_xu
	+\frac{2q_2\gamma^2}{\nu_2+\gamma^2}(v_d(\rho)-u) $$
we observe that it features interesting analogies with the second order macroscopic traffic model with driver-assist vehicles proposed in~\cite{delis2015CMA}. Specifically, on the right-hand side the first term proportional to $\partial_xu$ reminds of ACC vehicles while the second term proportional to $v_d(\rho)-u$ reminds of CACC vehicles as they have been heuristically modelled in~\cite{delis2015CMA} and discussed in the previous sections.

\section{Numerical experiments}
\label{sect:numerics}
In this section, we focus on the numerical description of the models introduced so far. We start from an analysis of the system~\eqref{eq:AR} with pressure field satisfying the relation $p'(\rho)=\gamma H\lambda(\rho)/2$. In the following, we refer to it as the kinetic ARZ model. For this model, we study the role of the headway $H$ and that of the sensitivity/reactiveness of the drivers $\lambda(\rho)$. Next, we analyse the differences between a conservative and a non-conservative numerical discretisation when different Riemann problems have to be studied. Once the kinetic ARZ model is validated, we explore the two proposed control strategies and their impact on the solution. Namely the binary variance approach detailed in Section~\ref{sect:bin_var} and the desired speed control explained in Section~\ref{sect:vd}. For these two situations, we explore the role played by the different macroscopic parameters originating from our kinetic derivation and the capability of the model to drive the solution towards some desired states. 

\subsection{Test 1: Finite Volume methods for the kinetic ARZ model}
\label{sect:test1}
Let us first give some insights into the necessary methods for numerical investigations of the derived controlled models. 

We focus first on the classical ARZ model~\eqref{eq:AR} which can be rewritten in a conservative form as follows
\begin{equation}
	\begin{cases}
		\partial_t\rho+\partial_x(\rho u)=0 \\
		\partial_ty+\partial_x(yu)=0,
	\end{cases}
	\label{eq:ARC}
\end{equation}
where the new variable $y$ is defined as $y:=\rho u+\rho p(\rho)$ and where $p(\rho)$ is a primitive of $p'(\rho)=\gamma H\lambda(\rho)/2$. We then use a fifth order WENO method combined with a Rusanov flux for treating the hyperbolic derivatives~\cite{shu}. Thus, given the flux function $F(U)=(\rho u,\,yu)^T$ with $U=(\rho,\,y)^T$, we first reconstruct the unknown values $U^-$, $U^+$ at the interfaces and then we employ the numerical Rusanov flux defined as:
\begin{align}\label{rus}
	& \mathcal{F}(U^-,\,U^+):=\frac{1}{2}\left[F(U^+)+F(U^-)-\Theta(A)(U^+-U^-)\right], \\
	& \Theta(A):=\max_{U\in[U^-,\,U^+]}\abs{\lambda(A(U))},\label{rus2}
\end{align}
where $\max_{U\in[U^-,\,U^+]}\abs{\lambda(A(U))}$ is the maximum modulus of the eigenvalues of the Jacobian matrix 
$$ A(U):=
\begin{pmatrix}
	u & \rho \\
	0 & -u-\frac{\gamma H}{2}\rho\lambda(\rho)
\end{pmatrix}. $$
The reconstruction of $\rho$, $y$ at the grid interfaces is performed as in the usual WENO framework~\cite{shu}. Finally, for the time derivative we use a second order Runge-Kutta explicit time discretisation. In particular, the time step $\Delta{t}$ is chosen according to the stability condition $\Delta{t}=0.5\Delta{x}/\max_{x\in\Omega}\{\mu_+,\,\mu_-\}$, where $\mu_{\pm}$ are the eigenvalues of the Jacobian matrix $A(U)$ of the flux.
In the following we will refer to this scheme as FV conservative scheme for the ARZ  model. 

An alternative numerical scheme for model~\eqref{eq:AR} consists in first writing the model as
$$	\begin{cases}
		\partial_t\rho+\partial_x(\rho u)=0 \\
		\partial_t(\rho u)+\partial_x(\rho u^2)=\rho^2\dfrac{\gamma H\lambda(\rho)}{2}\partial_x u
	\end{cases} $$
and then using the same scheme~\eqref{rus}-\eqref{rus2} with $U=(\rho,\,q)^T$, $q=\rho u$, and $F(U)=(q,\,q/\rho)^T$ and eigenvalues $\max\{u\in[u^-,\,u^+]\}$ with $u^-$ and $u^+$ the reconstructed velocities at the interfaces, the reconstruction of the macroscopic parameters on the grid interfaces following the same lines outlined in the previous paragraph. Then, the term
$$ \rho^2\frac{\gamma\lambda(\rho)H}{2}\partial_xu $$
appearing on the right-hand side of the momentum equation is treated by central difference discretisation. This leads to the following first order in time scheme
\begin{align*}
	&\rho^{n+1}_i=\rho^n_i-\frac{\Delta{t}}{\Delta{x}}\left(\hat{f}^n_{i+\frac{1}{2}}-\hat{f}^n_{i-\frac{1}{2}}\right), \\
	&q^{n+1}_i=q^n_i-\frac{\Delta{t}}{\Delta{x}}\left(\hat{g}^n_{i+\frac{1}{2}}-\hat{g}^n_{i-\frac{1}{2}}\right)-\Delta{t}{(\rho^{n}_i)}^2\frac{\gamma\lambda(\rho^n_i)H}{2}
		\cdot\frac{u^{n}_{i+1}-u^{n}_i}{\Delta{x}},
\end{align*} 
where $i=1,\,\dots,\,N$ is the number of spatial cells used with uniform $\Delta x$ and with $\hat{f}$ and $\hat{g}$ the two components of the numerical flux $\mathcal{F}$ in~\eqref{rus}.
The extension to a second order in time Runge-Kutta is then straightforward and is not discussed. Finally, we fix the time step according to the stability condition $\Delta{t}=0.5\Delta{x}/\max_{x\in\Omega}\{\mu_+,\,\mu_-\}$, where $\mu_\pm$ are the eigenvalues. In the following we will refer to this scheme as the non-conservative FV scheme for the ARZ model.

We will see in the next part that the conservative and non-conservative discretisations give in many cases analogous results. However, there are situations in which they may produce sensible differences.

Let us now consider four different Riemann problems. These have been inspired by the analysis performed in the seminal paper~\cite{aw2000SIAP} by Aw and Rascle. There, the authors discuss the possible analytical solutions which their model can furnish in terms of elementary waves: shock, rarefaction or combinations of these two waves. For these four different problems, we compare the cases $H>0$ and $H=0$. Moreover, we consider the case in which the sensitivity function $\lambda(\rho)$ in~\eqref{eq:binary} is either constant, i.e. independent of $\rho$, or defined as $\lambda(\rho)=\rho$. As shown by~\eqref{eq:AR.p}, this determines different expressions of the traffic pressure and consequently of the simulation results. The number of cells is fixed to $2000$, the final time is $T=10$, Dirichlet boundary conditions are taken.
We prescribe the following initial conditions:
\begin{align*}
&\text{RP1:}
	&
	&\rho_0(x)=0.9,\ -5\leq x\leq 5,
	&
	&u_0(x)=
		\begin{cases}
			0.5 & \text{if } x<0 \\
			0.25 & \text{if } x\geq 0,
		\end{cases} \\
&\text{RP2:}
	&
	&\rho_0(x)=0.9,\ -5\leq x\leq 5,
	&
	&u_0(x)=
		\begin{cases}
			0.9 & \text{if } x<-2.5 \\
			0.59 & \text{if } x\geq -2.5,
		\end{cases} \\
&\text{RP3:}
	&
	&\rho_0(x)=0.5,\ -5\leq x\leq 5,
	&
	&u_0(x)=
		\begin{cases}
			0.5 & \text{if } x<-2.5 \\
			0.3125 & \text{if } x\geq -2.5,
		\end{cases} \\
&\text{RP4:}
	&
	&\rho_0(x)=
		\begin{cases}
			0.5 & \text{if } x<-2.5 \\
			10^{-5} & \text{if } x\geq -2.5,
		\end{cases}
	&
	&u_0(x)=
		\begin{cases}
			0.25 & \text{if } x<-2.5 \\
			0 & \text{if } x\geq -2.5.
		\end{cases}
\end{align*}
These initial conditions have been chosen in such a way that, defining with $u^L$ the left initial speed of the vehicles and with $u^R$ the right initial one, for problem RP1 one has $u^L>u^R$, for problem RP2 one has $u^L<u^R<u^L+p(\rho)$ and for problem RP3 one has $u^R>u^L+p(\rho)$. Finally, in problem RP4 the right density is set to a value close to zero while the right speed is set to zero.

\begin{figure}
\centering
\includegraphics[width=0.42\textwidth]{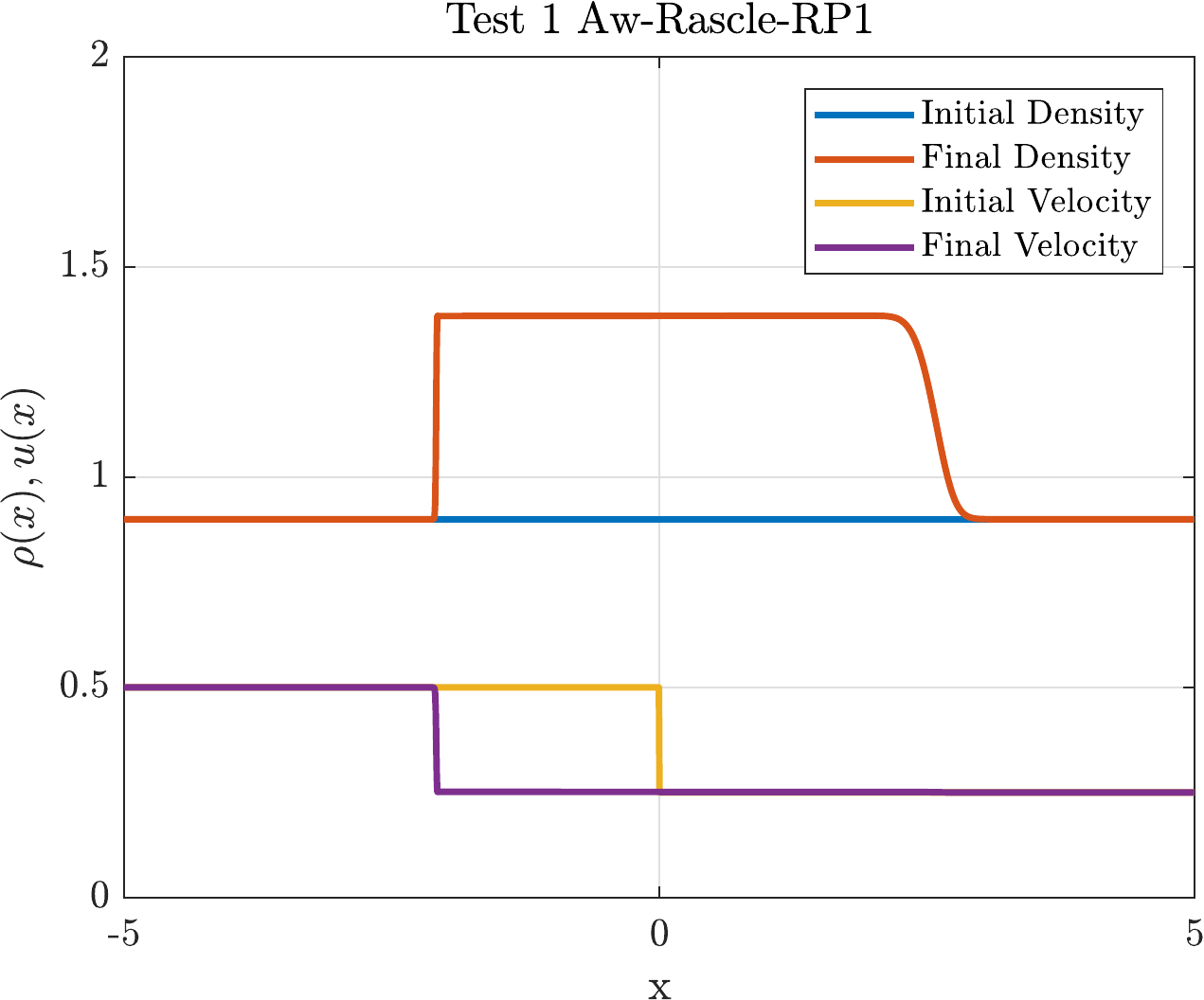}
\includegraphics[width=0.42\textwidth]{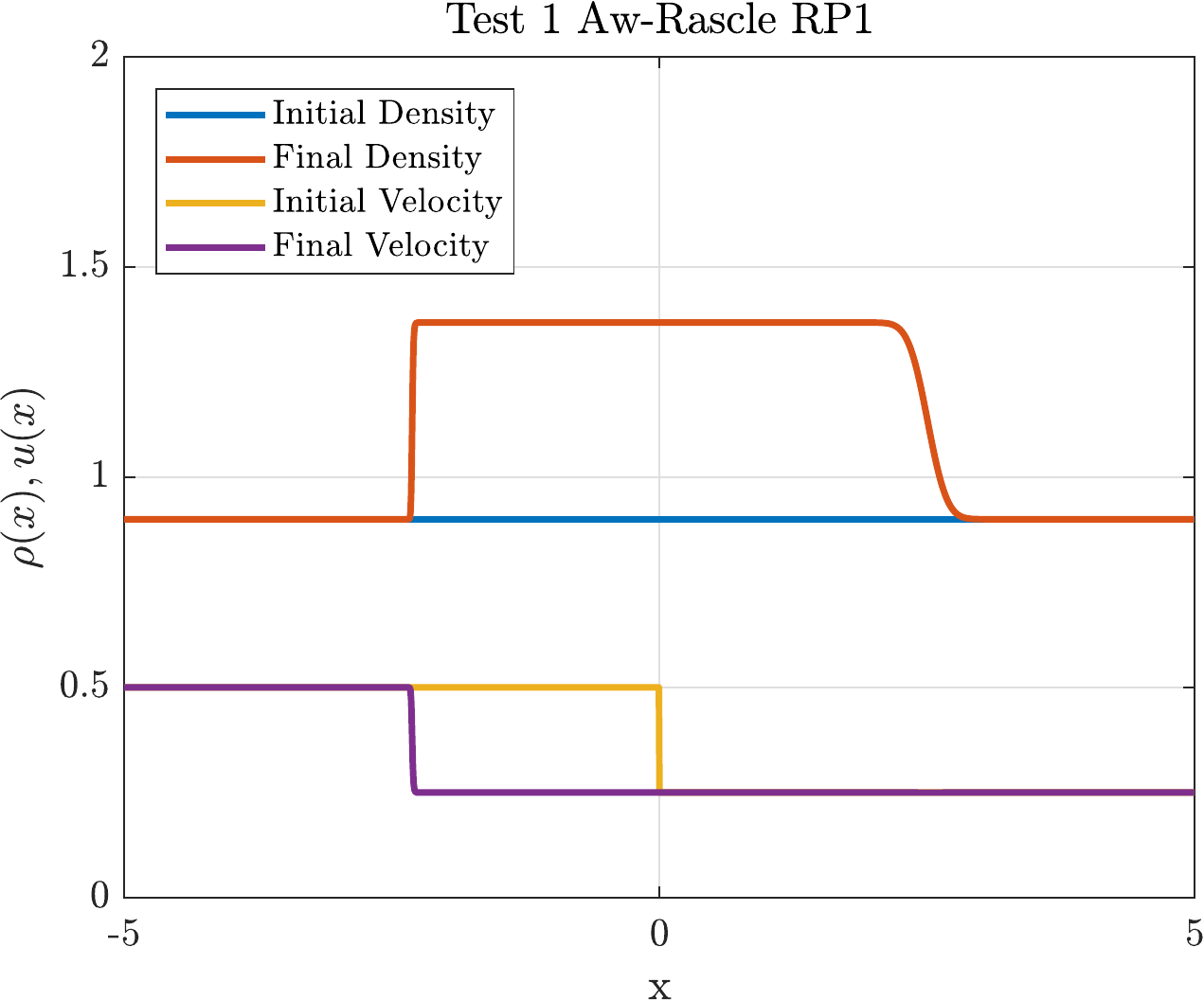} \\
\includegraphics[width=0.42\textwidth]{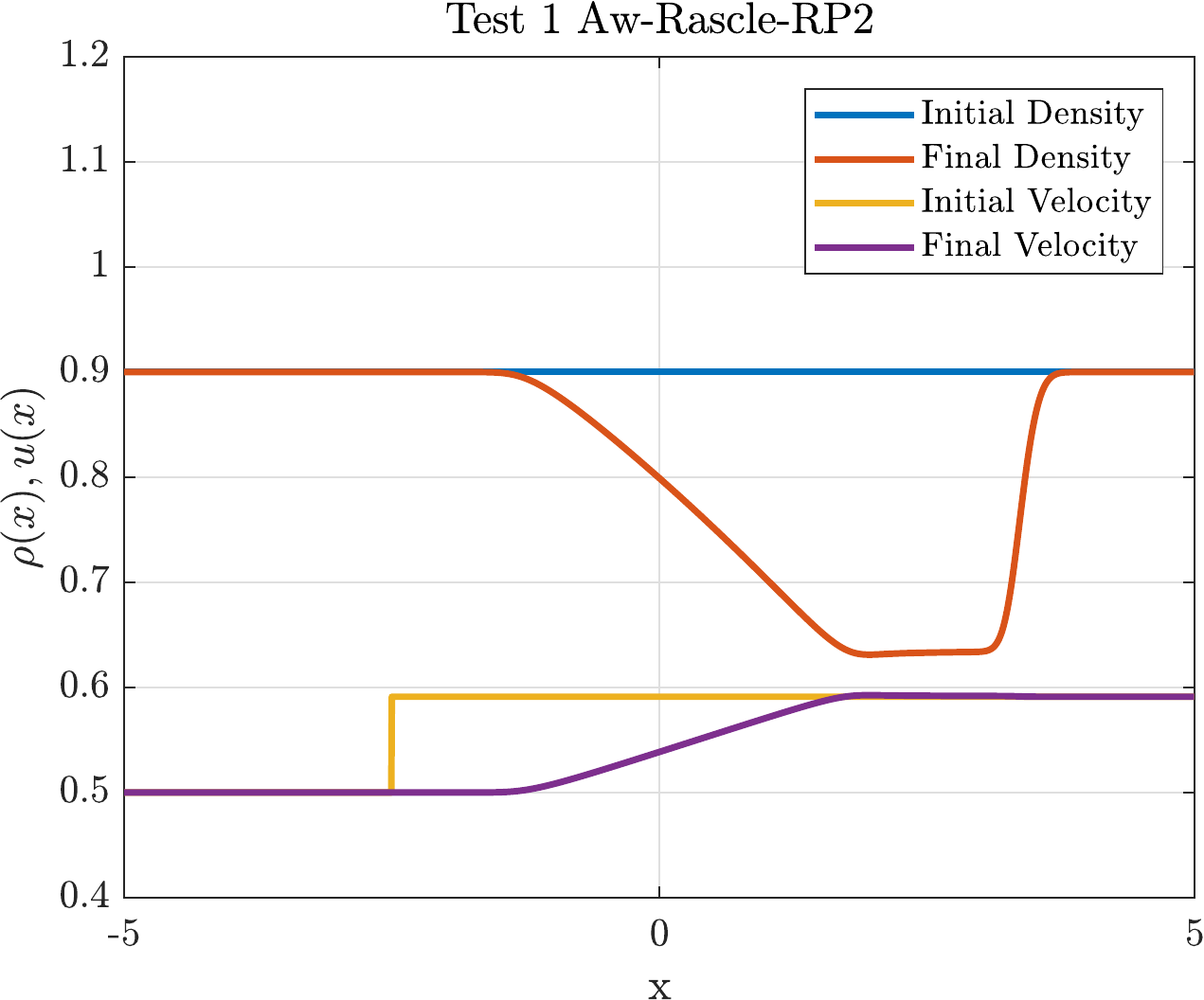}
\includegraphics[width=0.42\textwidth]{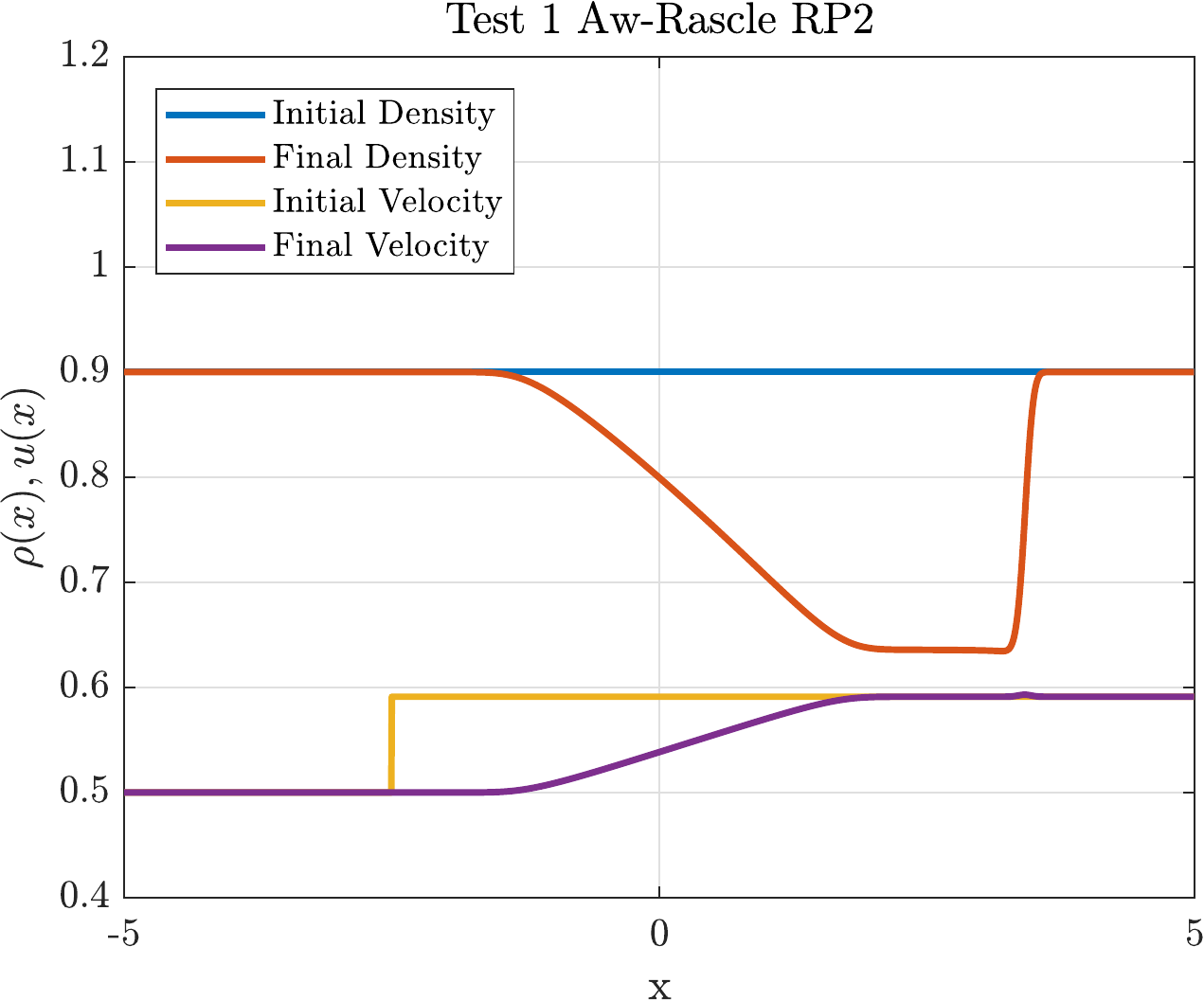} \\
\includegraphics[width=0.42\textwidth]{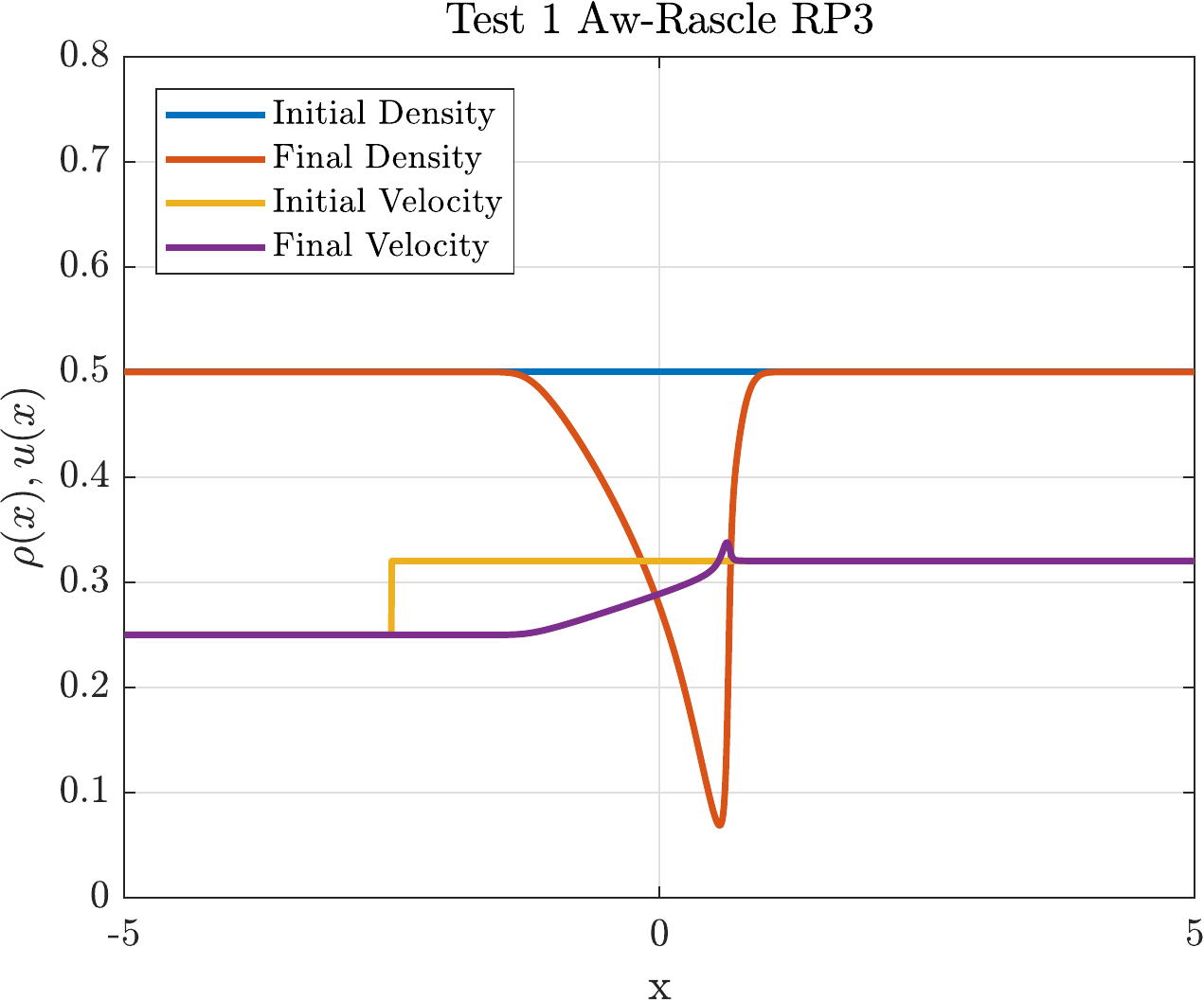}
\includegraphics[width=0.42\textwidth]{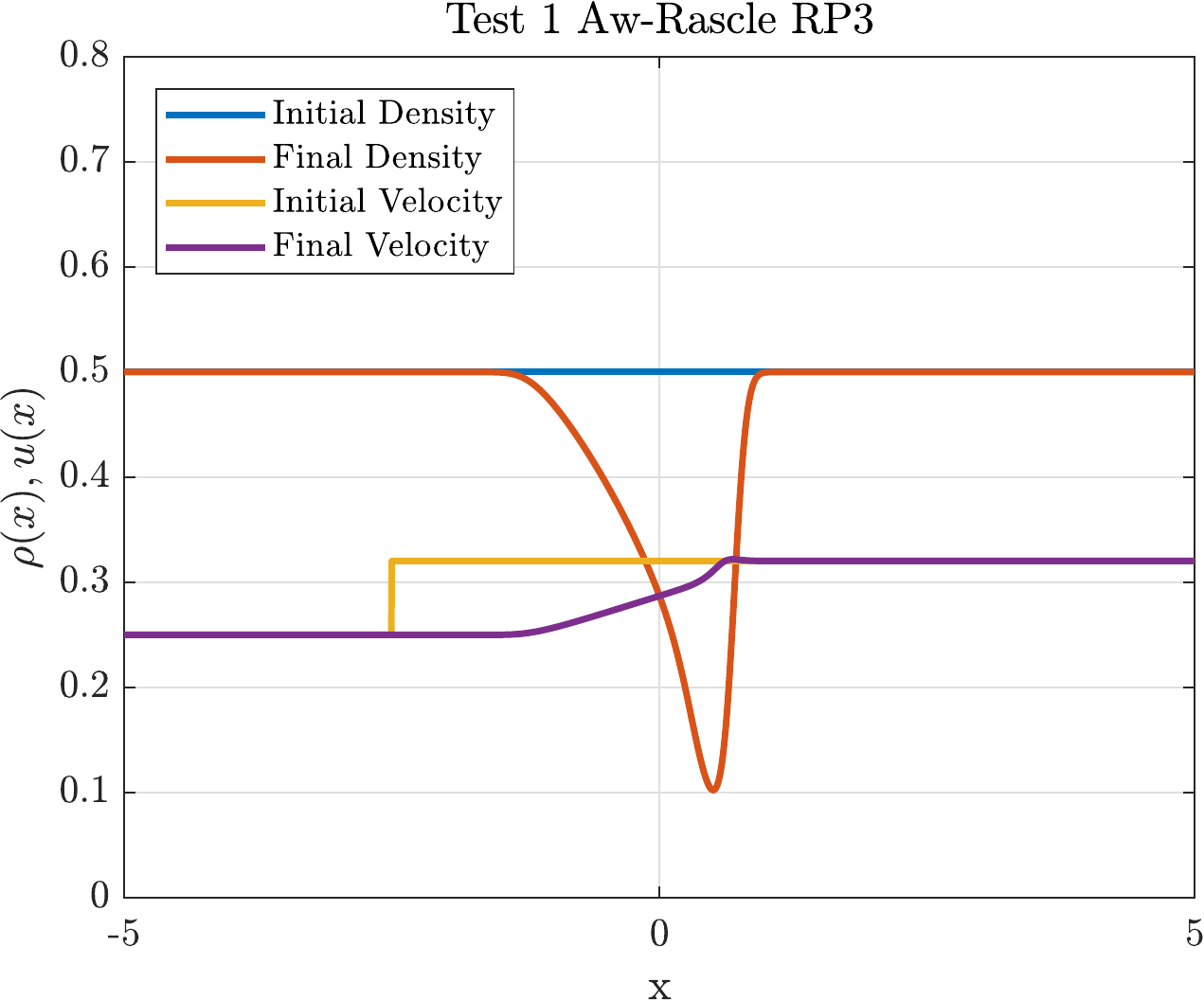} \\
\includegraphics[width=0.42\textwidth]{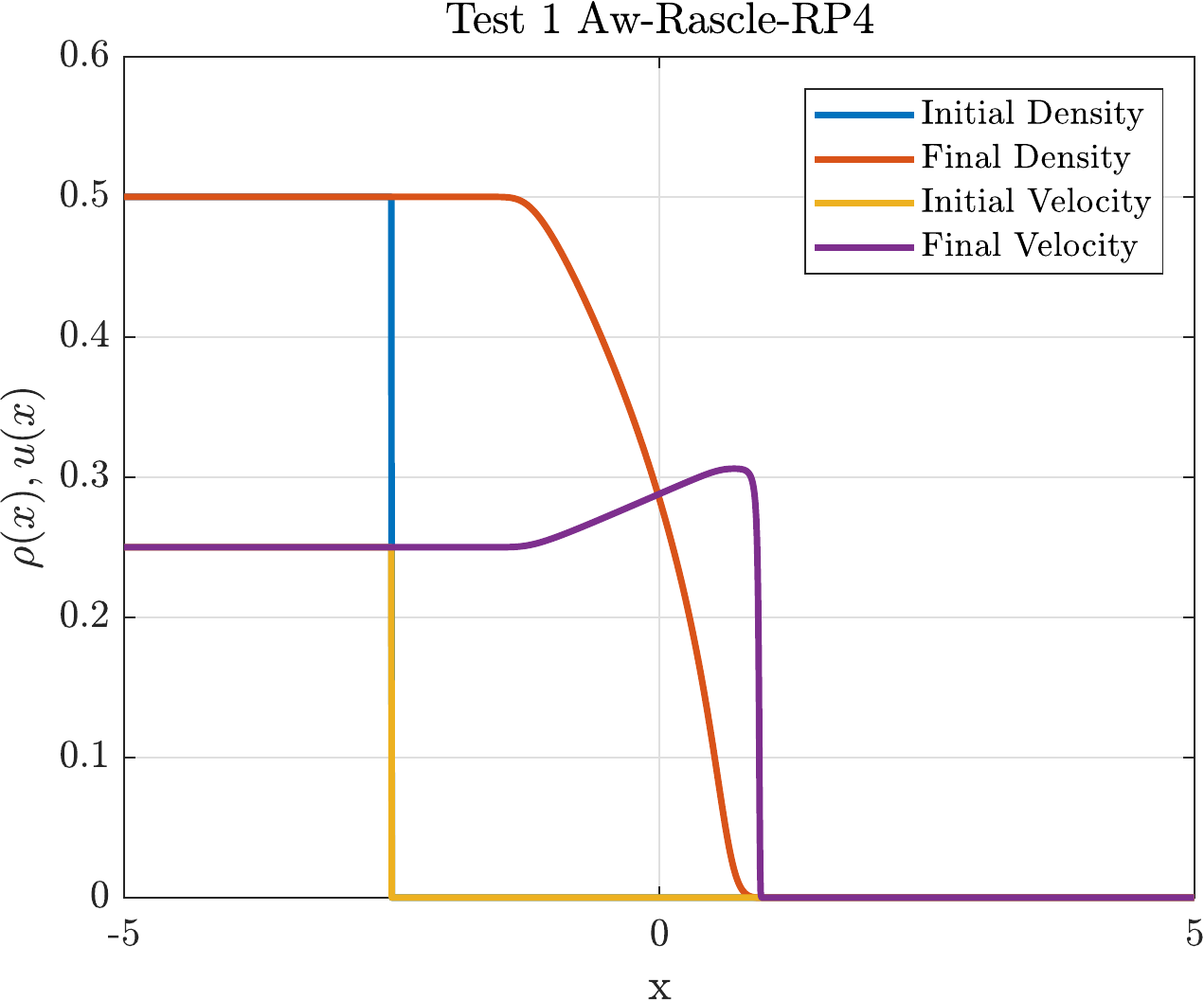}
\includegraphics[width=0.42\textwidth]{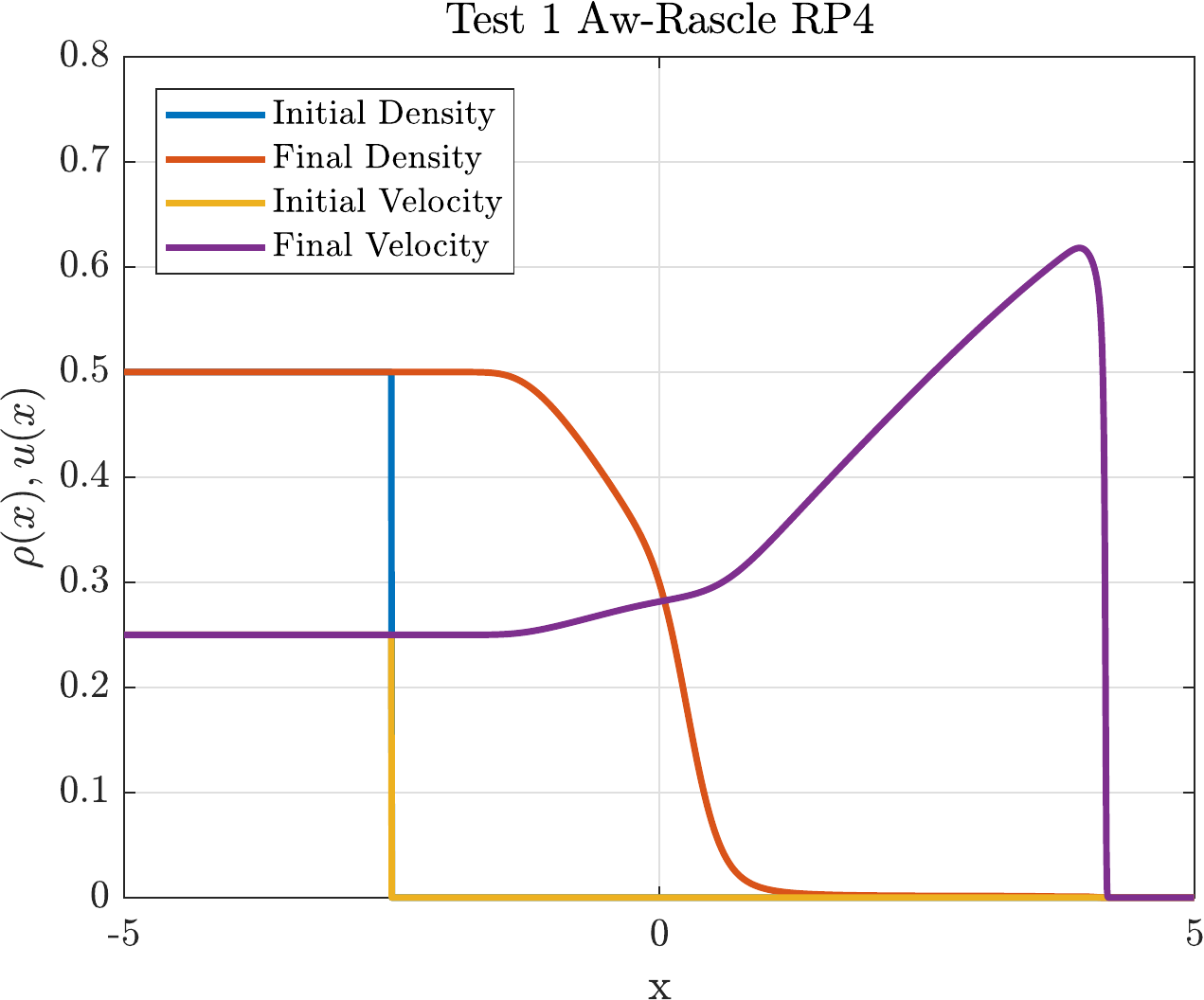}
\caption{\textbf{Test 1}. Four Riemann problems for the kinetic ARZ model. Initial and final states for density and speed of the vehicles with $\lambda(\rho)=\rho$. Left pictures: conservative scheme, right pictures: non-conservative scheme.}
\label{fig:test1}
\end{figure}

\begin{figure}[t]
\centering
\includegraphics[width=0.515\textwidth]{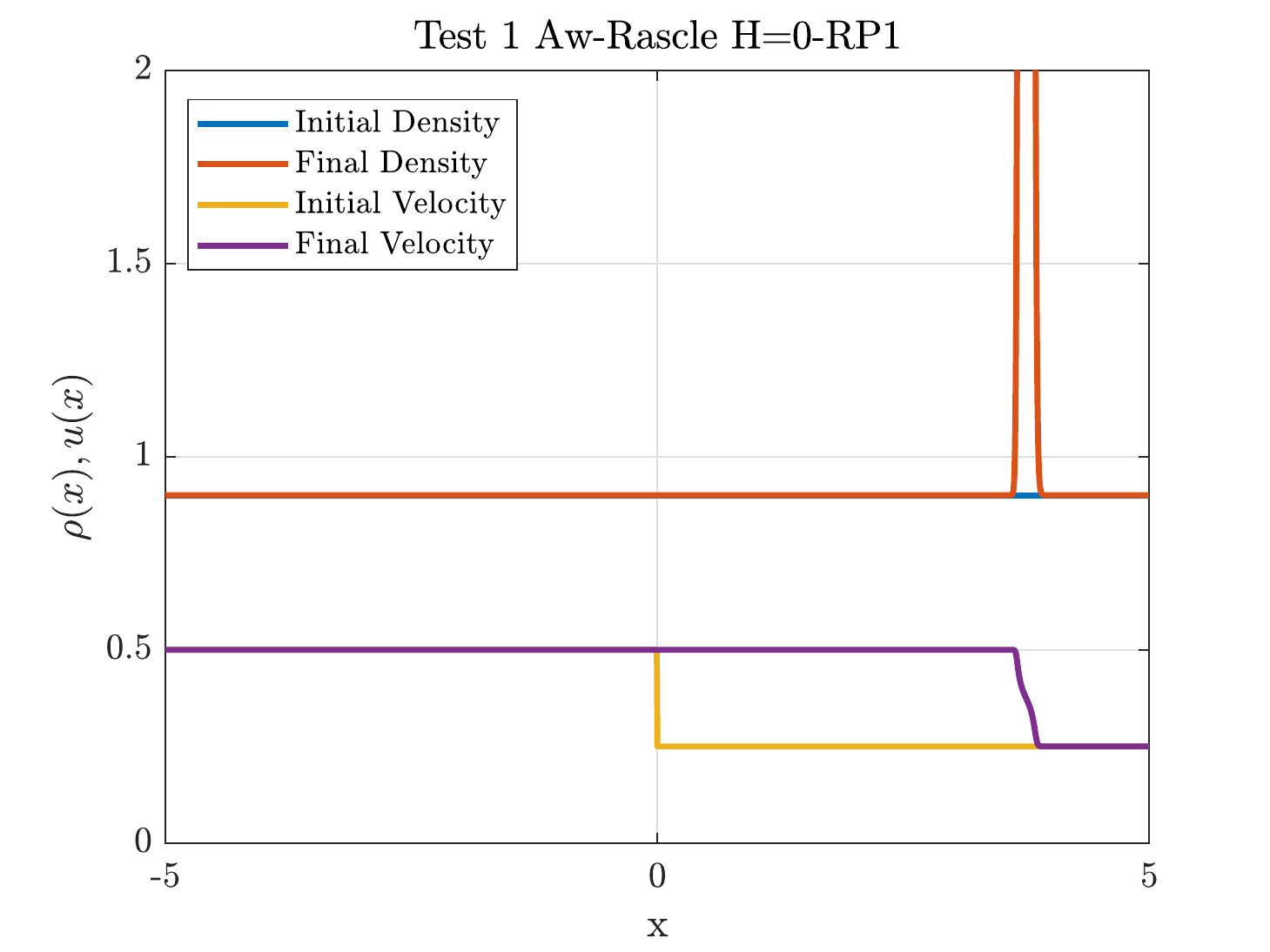}
\includegraphics[width=0.45\textwidth]{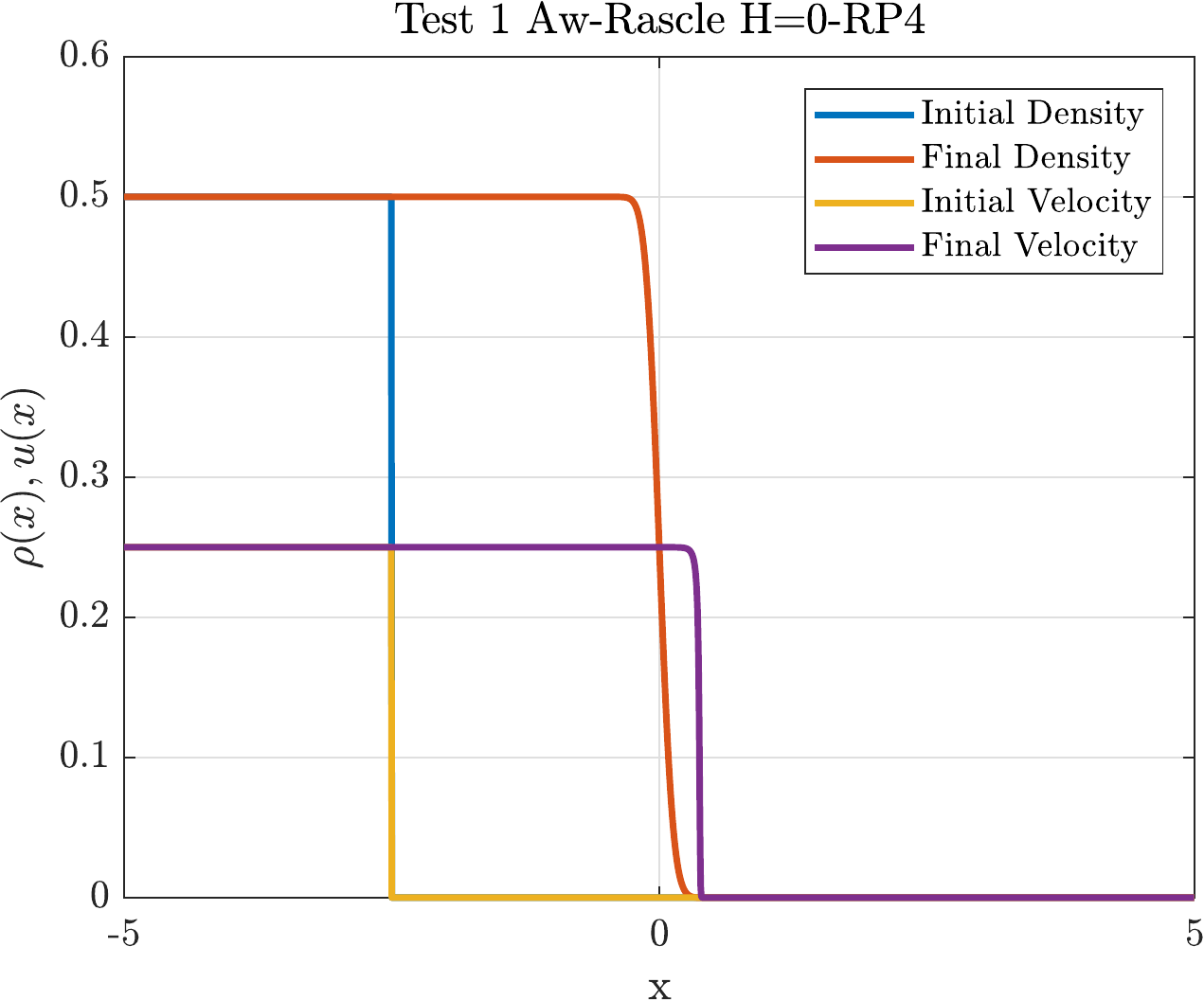}
\caption{\textbf{Test 1}. Riemann problems RP1 and RP4 for the kinetic ARZ model with $H=0$. Initial and final states for density and speed of the vehicles with $\lambda(\rho)=\rho$.}
\label{fig:test12}
\end{figure}

In Figure~\ref{fig:test1}, the initial and final density and speed of the vehicles are shown for the four Riemann problems in the case $H>0$ and $\lambda(\rho)=\rho$. On the left we show the results of the conservative scheme and on the right those of the non-conservative scheme. In Figure~\ref{fig:test12}, the same quantities are shown for the first and fourth Riemann problems in the case $H=0$. In this case, the conservative and non-conservative schemes lead to the same discretisation since $p(\rho)=0$. Moreover, we only show the results for problems RP1 and RP4 since for problems RP2 and RP3 the density and speed are constant in time and coincide with their respective initial data. This is due to the fact that $H=0$ implies $u^L=u^R$.

\begin{figure}[t]
\centering
\includegraphics[width=0.45\textwidth]{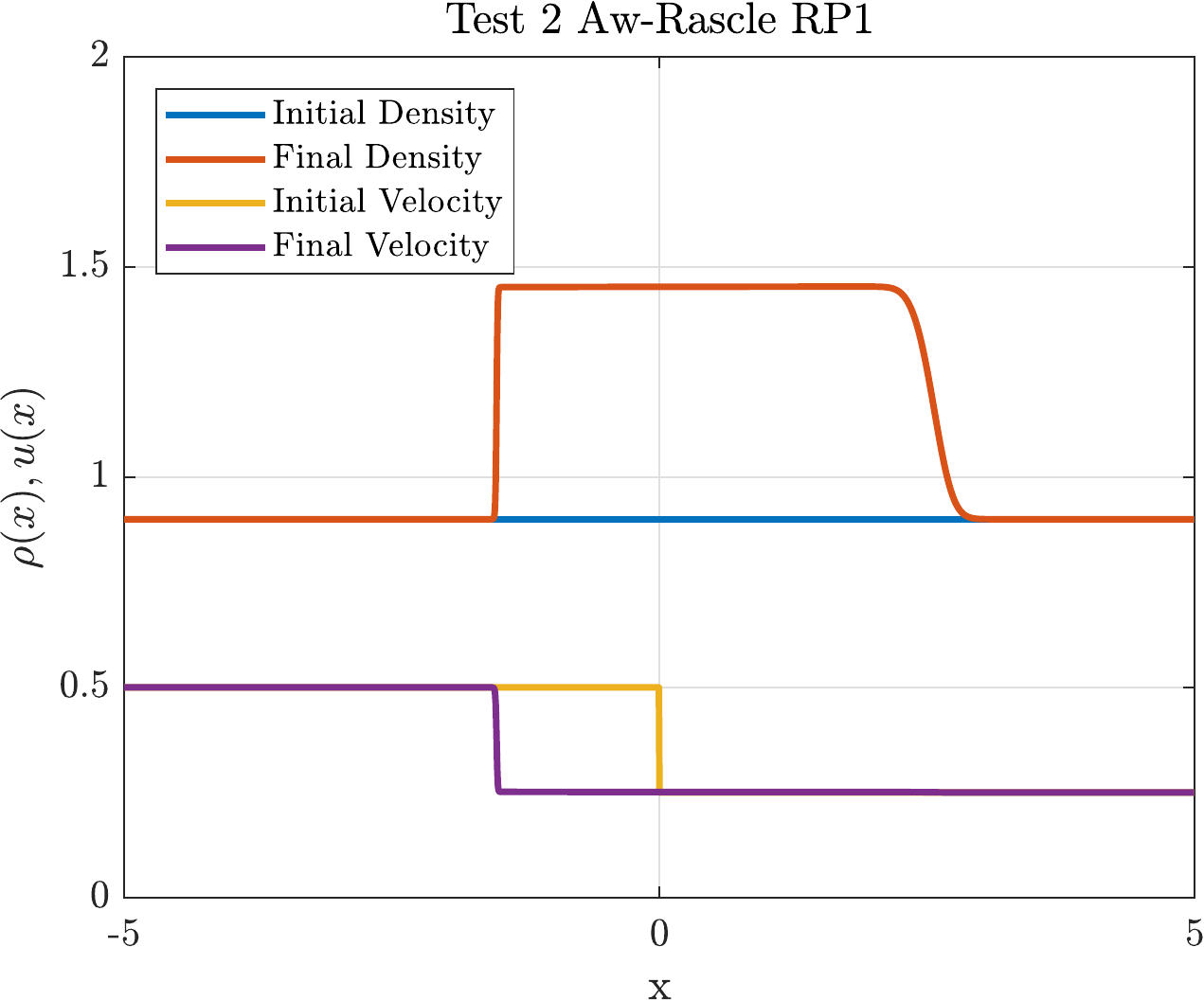}
\includegraphics[width=0.45\textwidth]{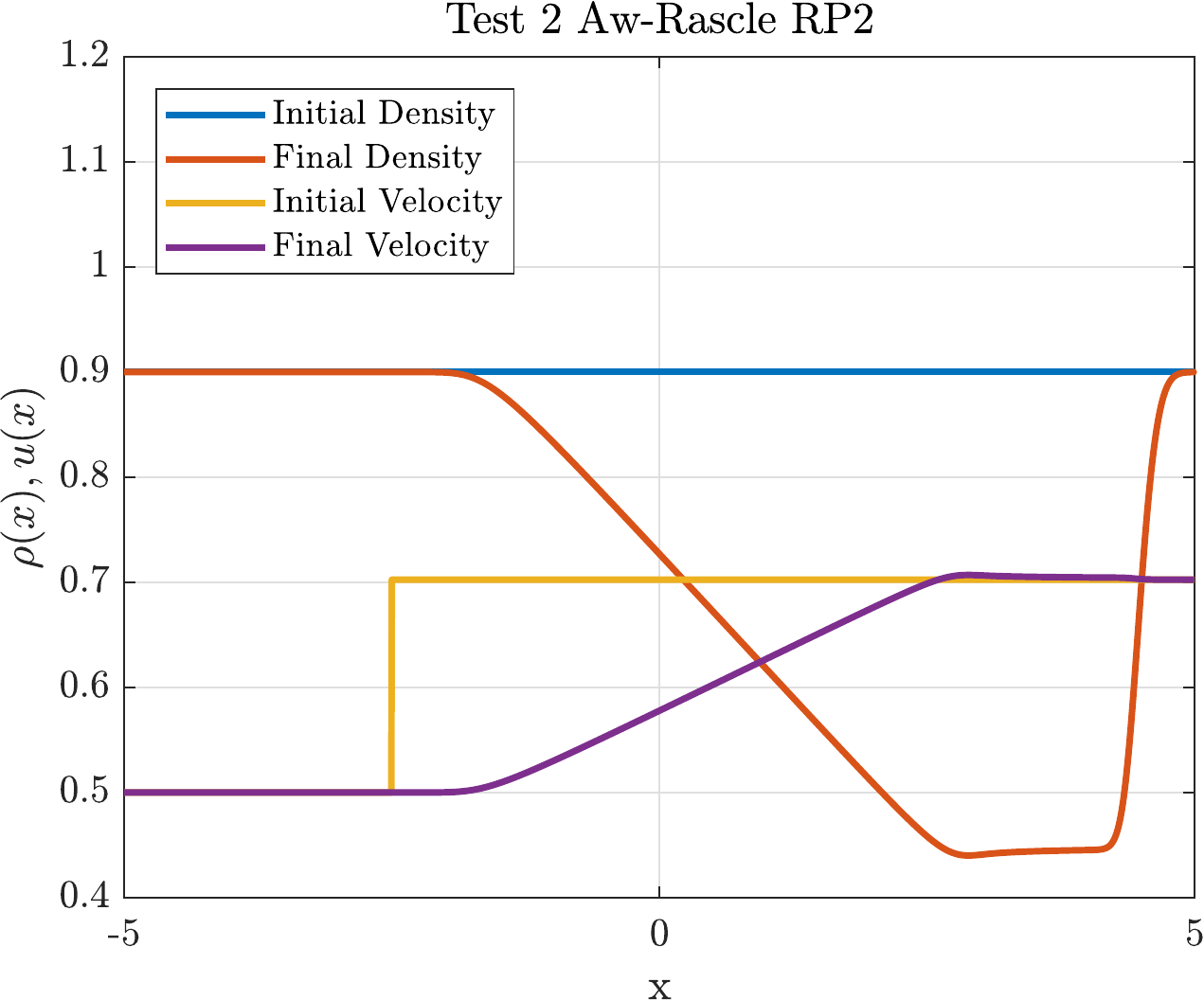} \\
\includegraphics[width=0.45\textwidth]{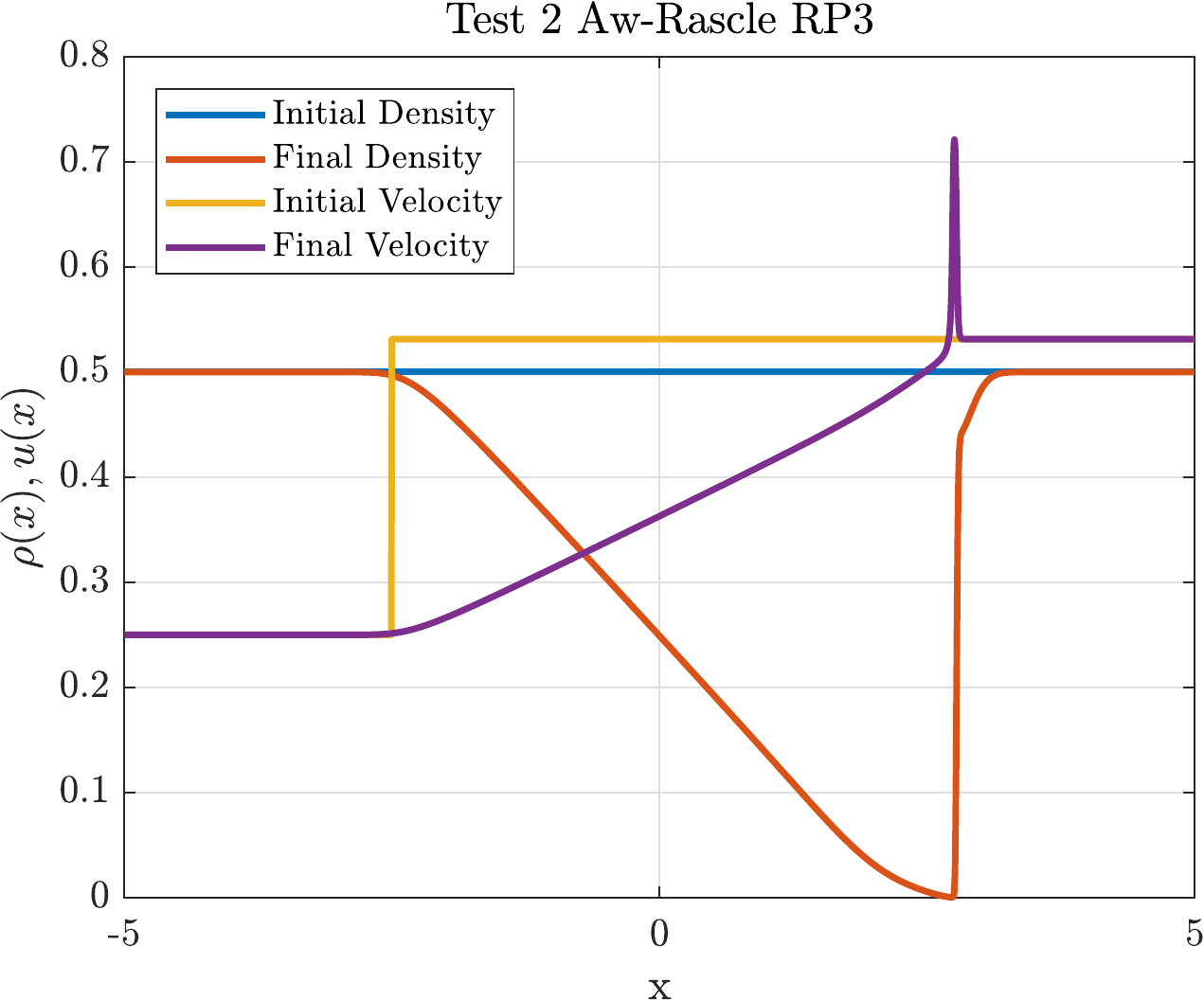}
\includegraphics[width=0.45\textwidth]{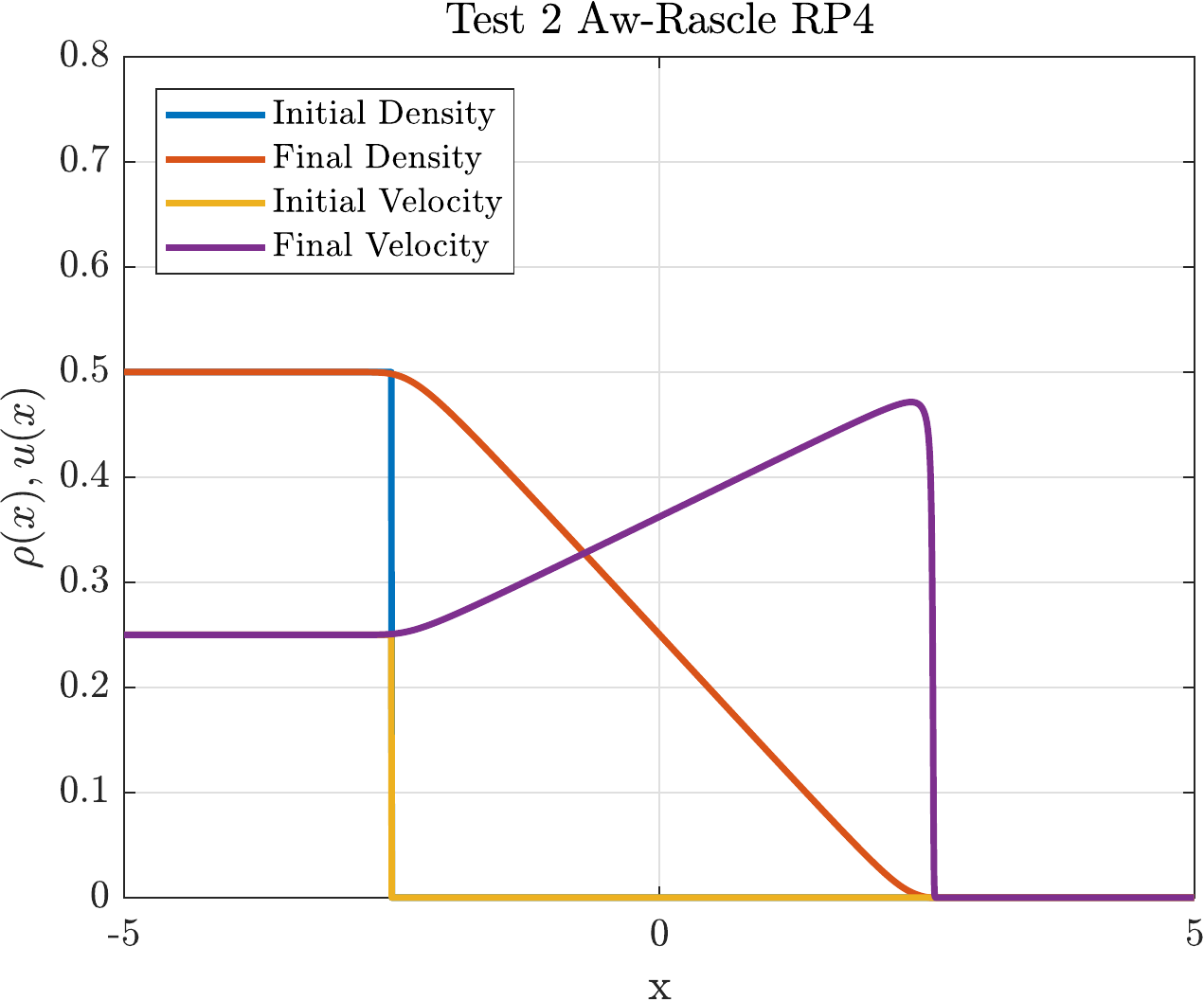}
\caption{\textbf{Test 1}. Four Riemann problems for the kinetic ARZ model and conservative discretisation. Initial and final states for density and speed of the vehicles with $\lambda(\rho)=1$.}
\label{fig:test2}
\end{figure}

In the case $H>0$, for RP1 the solution is a shock wave moving to the left for both density and speed combined with another smoother shock wave moving to the right for the sole density of vehicles. For RP2 we have a first rarefaction wave for both macroscopic quantities followed by a second rarefaction for the density. For RP3 we have first a rarefaction followed by a shock for both density and speed. Finally, RP4 presents a smooth decrease in the density accompanied by a vehicles acceleration. To that situation, it follows a sudden jump to zero when the vacuum condition is reached. If one observes the trend of the same simulations when $H=0$ the differences are evident. For RP1 the shock moves to the right and it leads to a delta whereas when $H>0$ the traffic jam has lower intensity and realistically it moves backwards. The case $H=0$ gives for this problem unrealistic results. The cases RP2 and RP3 cannot be reproduced when $H=0$, highlighting the fact that a non-zero headway leads to a much richer set of possible solutions. Finally, the case RP4 is also different, indeed for $H=0$ vehicles do not accelerate in presence of vacuum which instead is the typical observed driver reaction in these situations if the maximum allowed speed is not yet reached. Concerning the differences between the conservative and the non-conservative discretisations, one can appreciate macroscopic differences only for RP4 initial data, i.e. in the case of vacuum. For the other three problems results are nearly identical. The differences in RP4 are due to the difficulties in discretising the derivative $\partial_x u$ near to vacuum in the non-conservative case. Indeed, this scheme requires much smaller time steps to be stable while results seems to be largely affected by this choice. For this reason, in the following we restrict our results to the case of the sole conservative discretisation. To that aim, we show a last result for this section in Figure~\ref{fig:test2}. These pictures refer to the case in which the sensitivity $\lambda$ is independent of the vehicle density. In this case, the solutions suggest that vehicles move rightwards at higher pace, being all waves shifted in that direction compared to the case $\lambda(\rho)=\rho$. This is natural since the effects of the headway $H$ are less significant due to the fact that $\rho<1$ at least for problems RP2, RP3 and RP4. Problem RP1 is different and one can observe that the extension of the traffic jam is smaller when the sensitivity function is $\lambda(\rho)=1$. This is due to the fact that the pressure wave is slowing down the second shock wave appearing on the right for the density of vehicles. 

\subsection{Test 2: Controlled ARZ model with binary control strategies}
In this section, we discuss the kinetic ARZ model with binary control introduced in Section~\ref{sect:bin_var}. The hydrodynamic model is the one given in equation~\eqref{eq:ARZ.controlled.bin_var}, which is first rewritten in conservative form like in~\eqref{eq:ARC}. The variable $y$ still expresses the sum of the momentum and of the pressure field:
$$ y=\rho u+\rho P, \qquad P'(\rho):=\frac{\gamma H}{2}\left(1+q\gamma\frac{1-\gamma\lambda(\rho)}{(\nu+\gamma^2)\lambda(\rho)}\right)\lambda(\rho), $$
where $\lambda(\rho)=\rho$ and now the pressure field is the sum of the traffic pressure and the \textit{control pressure}. The numerical method then follows the lines of the one described in Test~\ref{sect:test1}, i.e. it is a combination of WENO reconstruction with Rusanov fluxes for the space derivative with a second order explicit Runge-Kutta method in time. The initial data are given by
$$	\rho_0(x)=0.9,\ -20\leq x\leq 20,
	\qquad
	u_0(x)=
	\begin{cases}
		0.25 & \text{if } x<-2.5\ \text{or} \ x>2.5 \\
		0.1 & \text{if } -2.5\leq x\leq 2.5.
	\end{cases} $$
The domain $x\in[-20,20]$ is paved with $1000$ cells and the final time is fixed to $T=75$. The boundary conditions are of Neumann type.

\begin{figure}[t]
\centering
\includegraphics[width=0.45\textwidth]{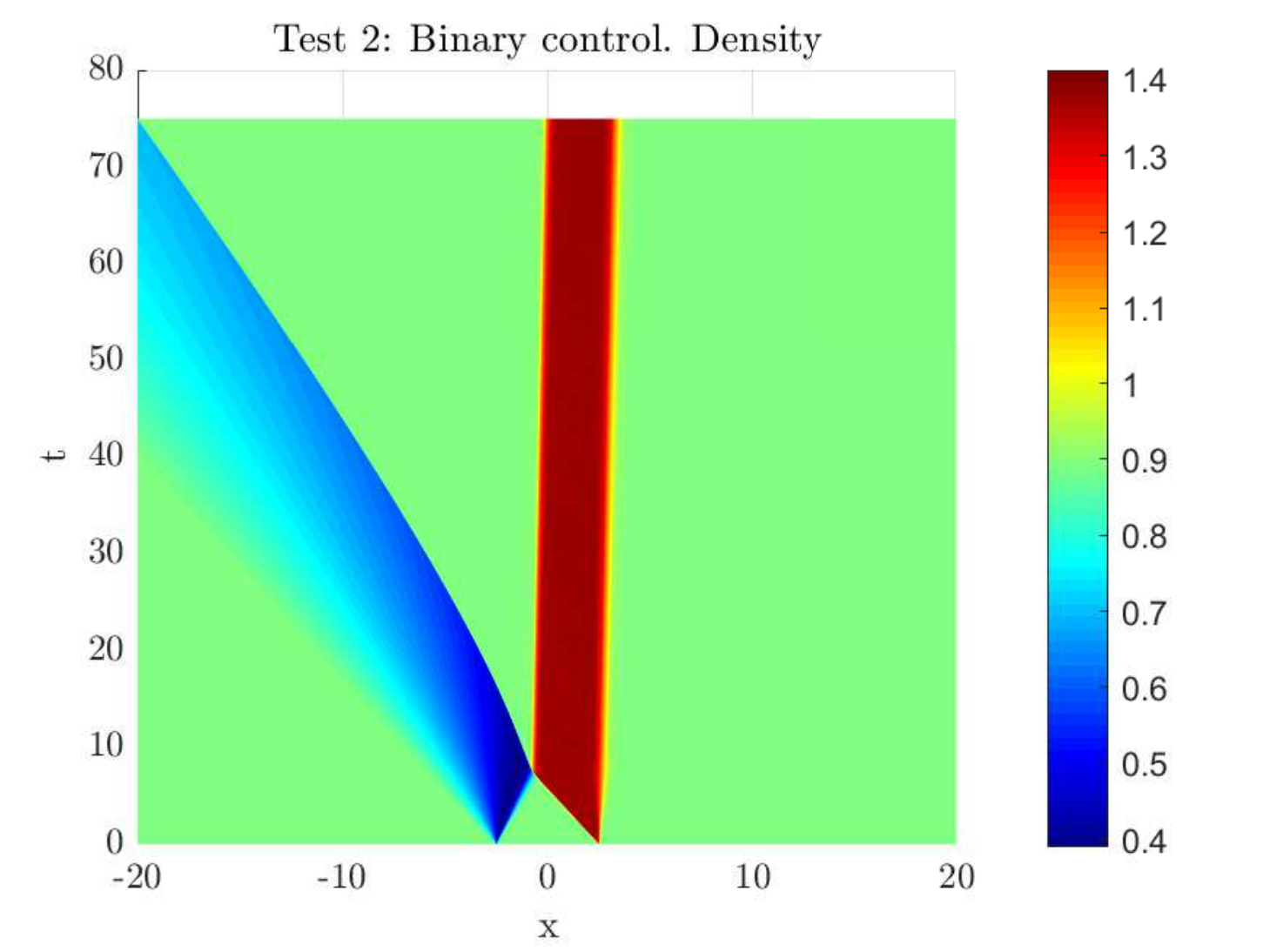}
\includegraphics[width=0.45\textwidth]{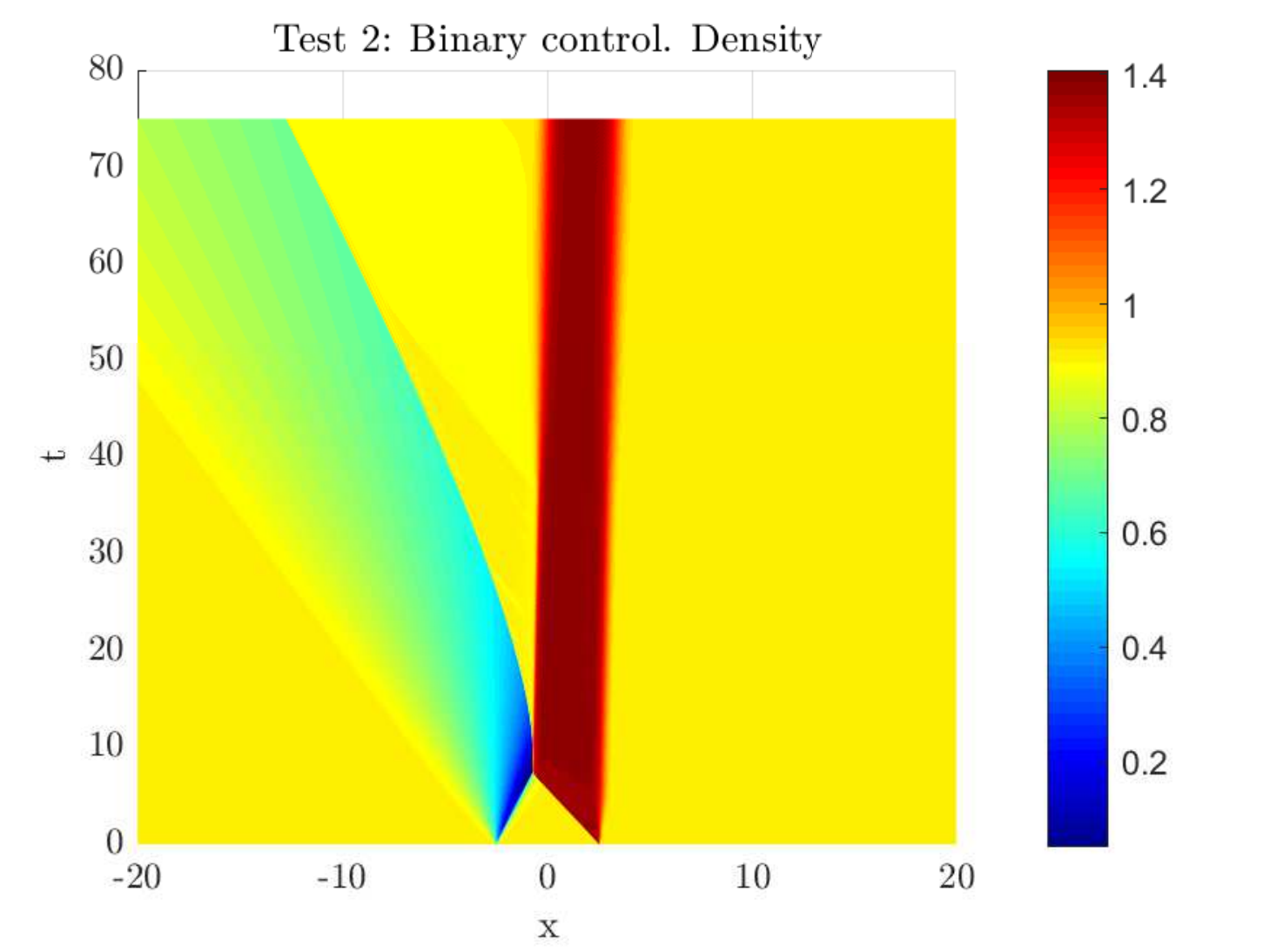} \\
\includegraphics[width=0.45\textwidth]{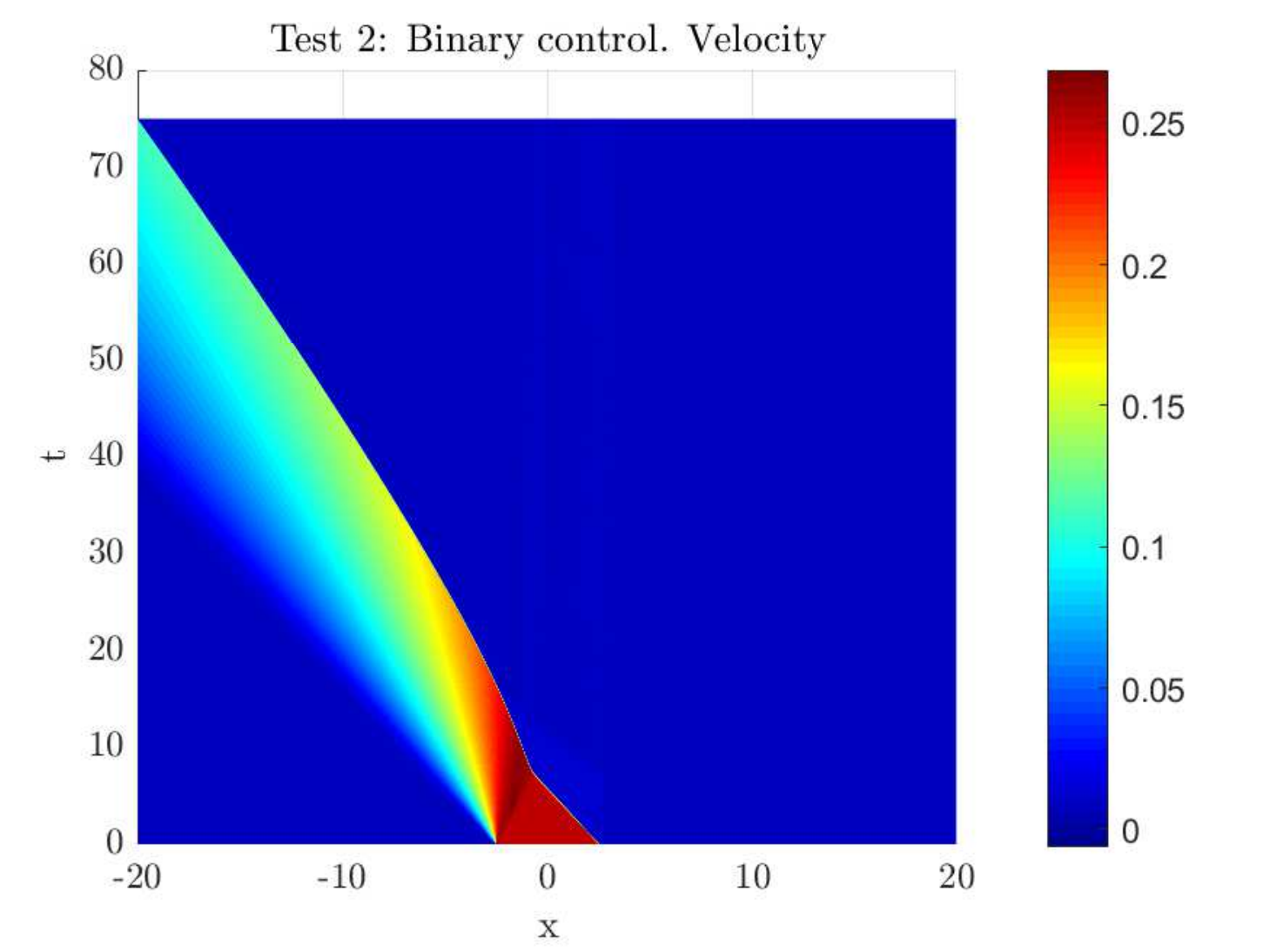}
\includegraphics[width=0.45\textwidth]{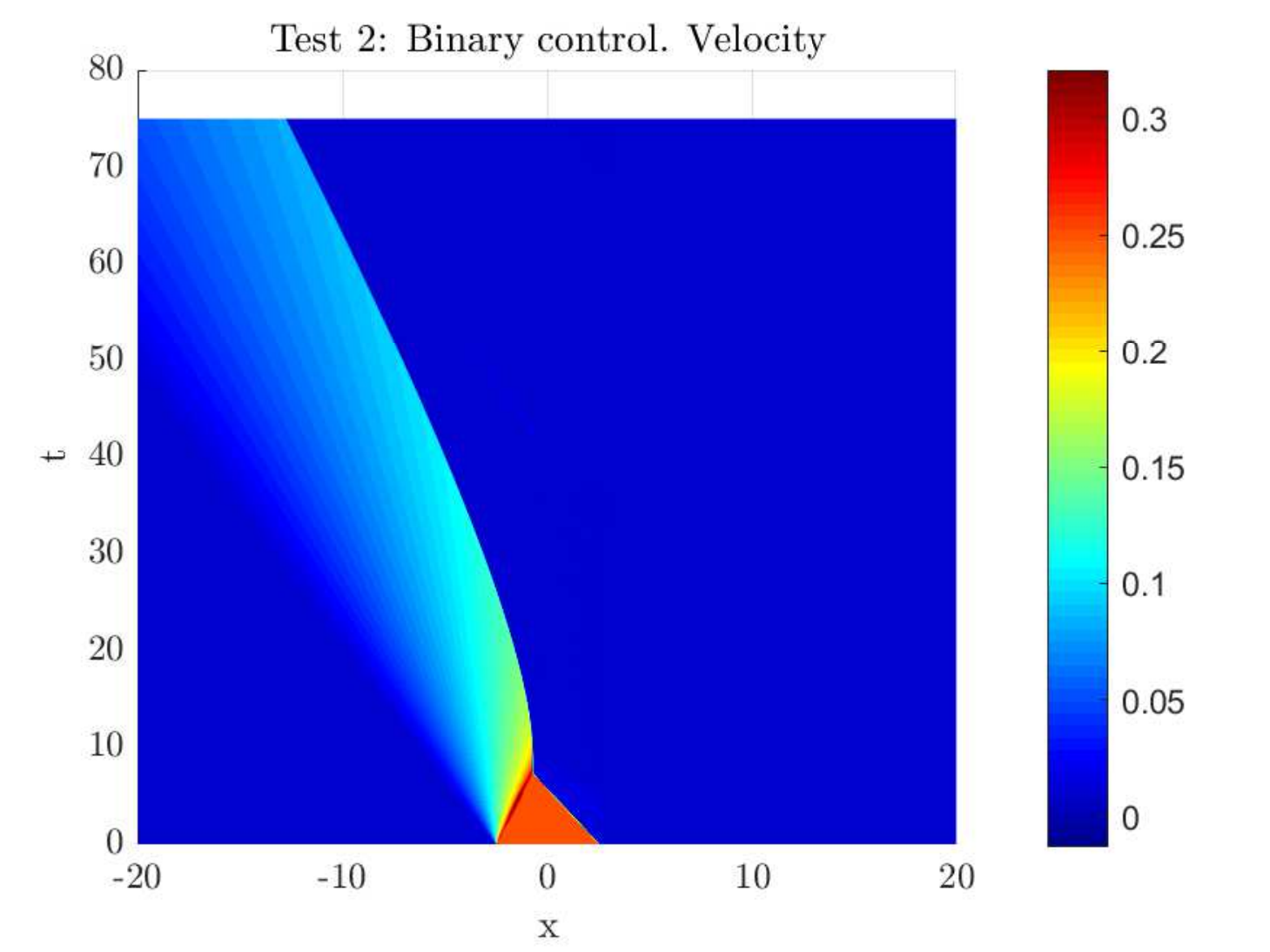}
\caption{\textbf{Test 2A}. Time evolution of the controlled ARZ model with binary control strategy. Top pictures: density, bottom pictures: mean speed with $\lambda(\rho)=\rho$. Left pictures show the case $q=1$, right pictures the case $q=0$.}
\label{fig:testc_binary1}
\end{figure}

In Figure~\ref{fig:testc_binary1} we show the evolution in the space-time domain of the density of vehicles together with their speed. Left pictures show the case $q=1$, right pictures the case $q=0$. The initial data are such that for the density of cars a steady shock wave forms in correspondence of $x=2.5$ while a shock followed by a leftwards moving rarefaction wave forms at $x=-2.5$. On the other hand, the mean speed consists only of a rarefaction followed by a shock moving leftwards at the same speed as the waves in the density field. At the end of the simulation, the waves have reached the left boundary and as expected the mean speed of the vehicles is constant. This has been obtained thanks to the pressure field generated by the microscopic binary control term, which determines a modification of the eigenvalues of the hydrodynamic model. At the end of the simulation, the steady shock wave in the density is instead still present. In more details, even for the case of $q=0$, i.e. the non-controlled case (right pictures), the system reaches an equilibrium mean speed. This is due to the presence of a pressure field which causes information to travel backwards like for the controlled case. The main difference between the controlled and the non-controlled cases is that in the case $q=1$ the constant mean speed is reached faster than in the case $q=0$ thanks to the control mechanism.

\begin{figure}[t!]
\centering
\includegraphics[width=0.45\textwidth]{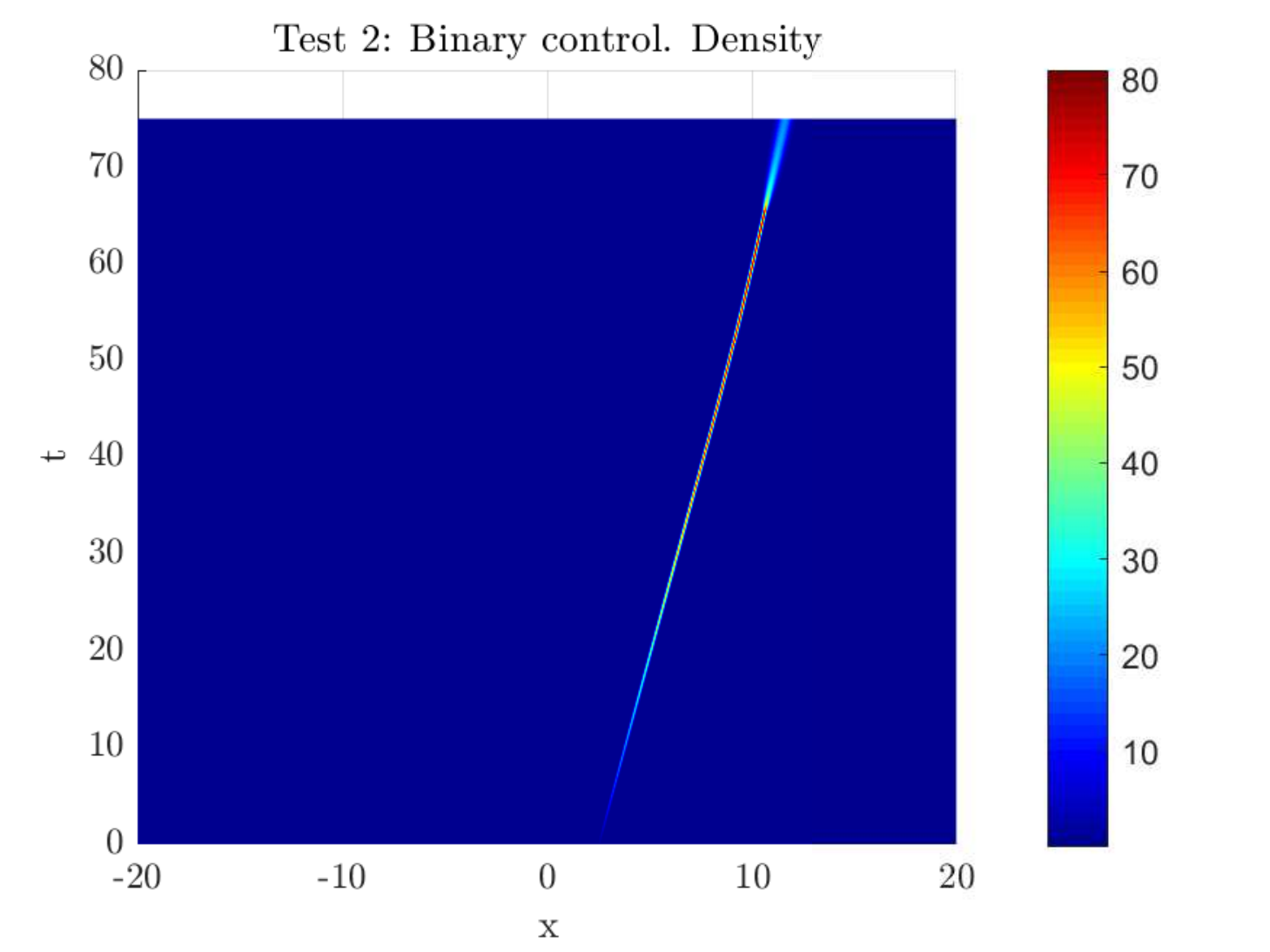}
\includegraphics[width=0.45\textwidth]{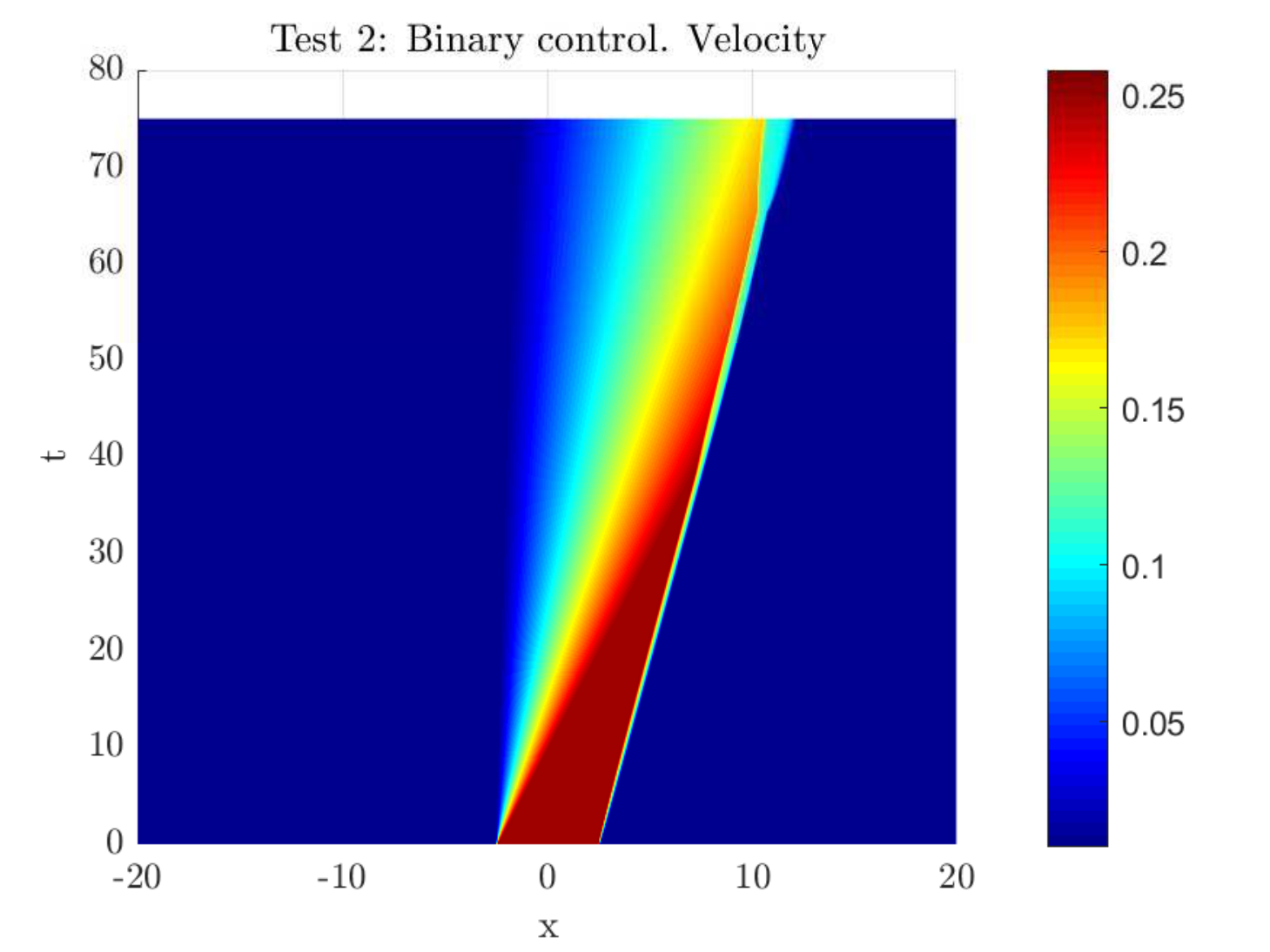} \\
\caption{\textbf{Test 2A}. Time evolution of the controlled ARZ model with the binary control strategy. Left picture: density, right picture: mean speed with $\lambda(\rho)=\rho$. Case $H=0$.}
\label{fig:testc_binary2}
\end{figure}

In Figure~\ref{fig:testc_binary2}, we report instead the case $H=0$. Here, the situation looks completely different. A delta shock in the density forms and moves rightwards, i.e. in the opposite direction with respect to the previous case. This is an unphysical and unobserved behaviour in real situations. Concerning the mean speed of the vehicles, we have that a rarefaction is followed by a shock both moving rightwards. In other words, there is no backward propagation of the information and vehicles move independently of what happens ahead. We stress that this does not represent the empirically observed evidences.

\begin{figure}[t]
\centering
\includegraphics[width=0.45\textwidth]{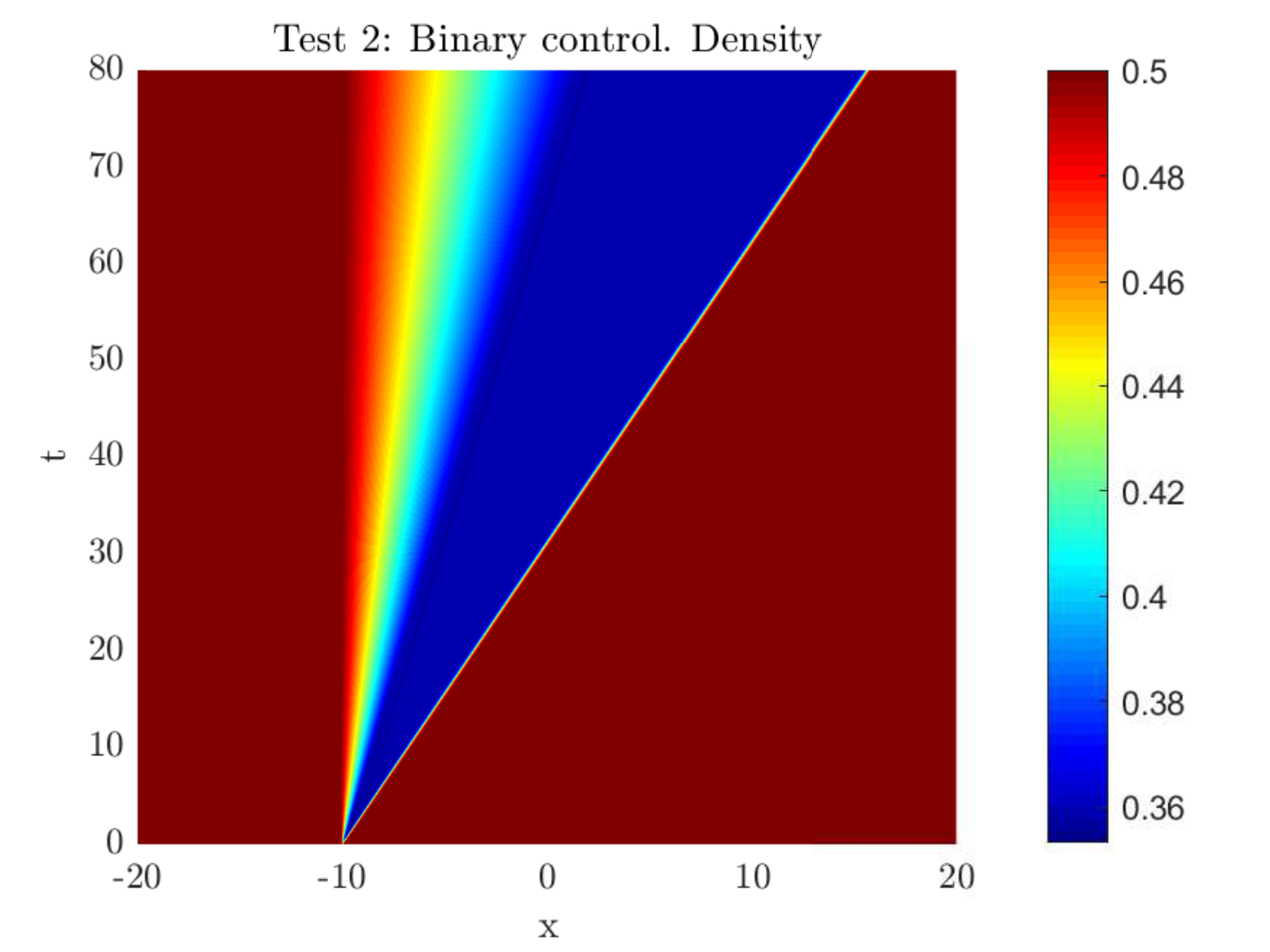}
\includegraphics[width=0.45\textwidth]{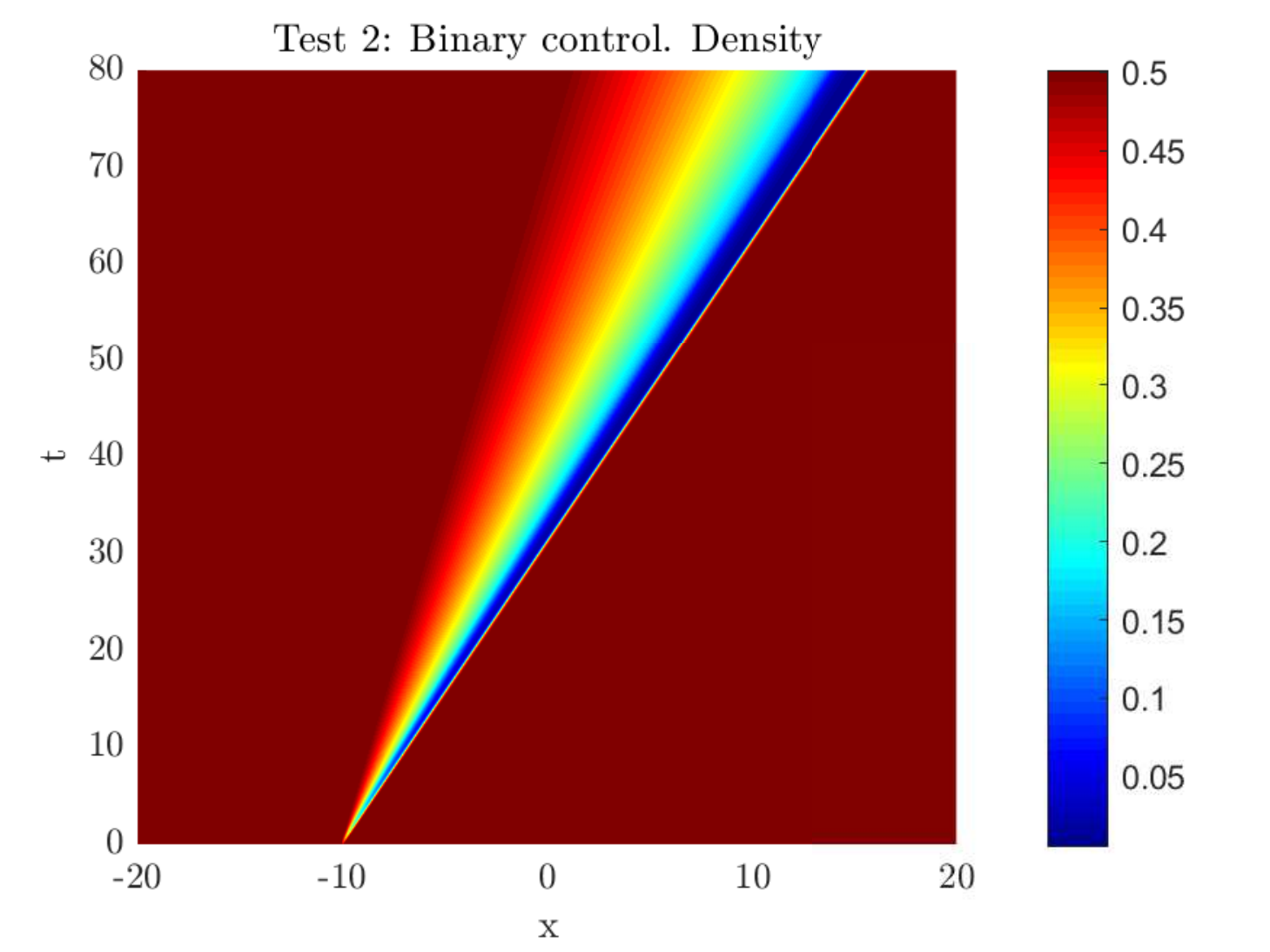} \\
\includegraphics[width=0.45\textwidth]{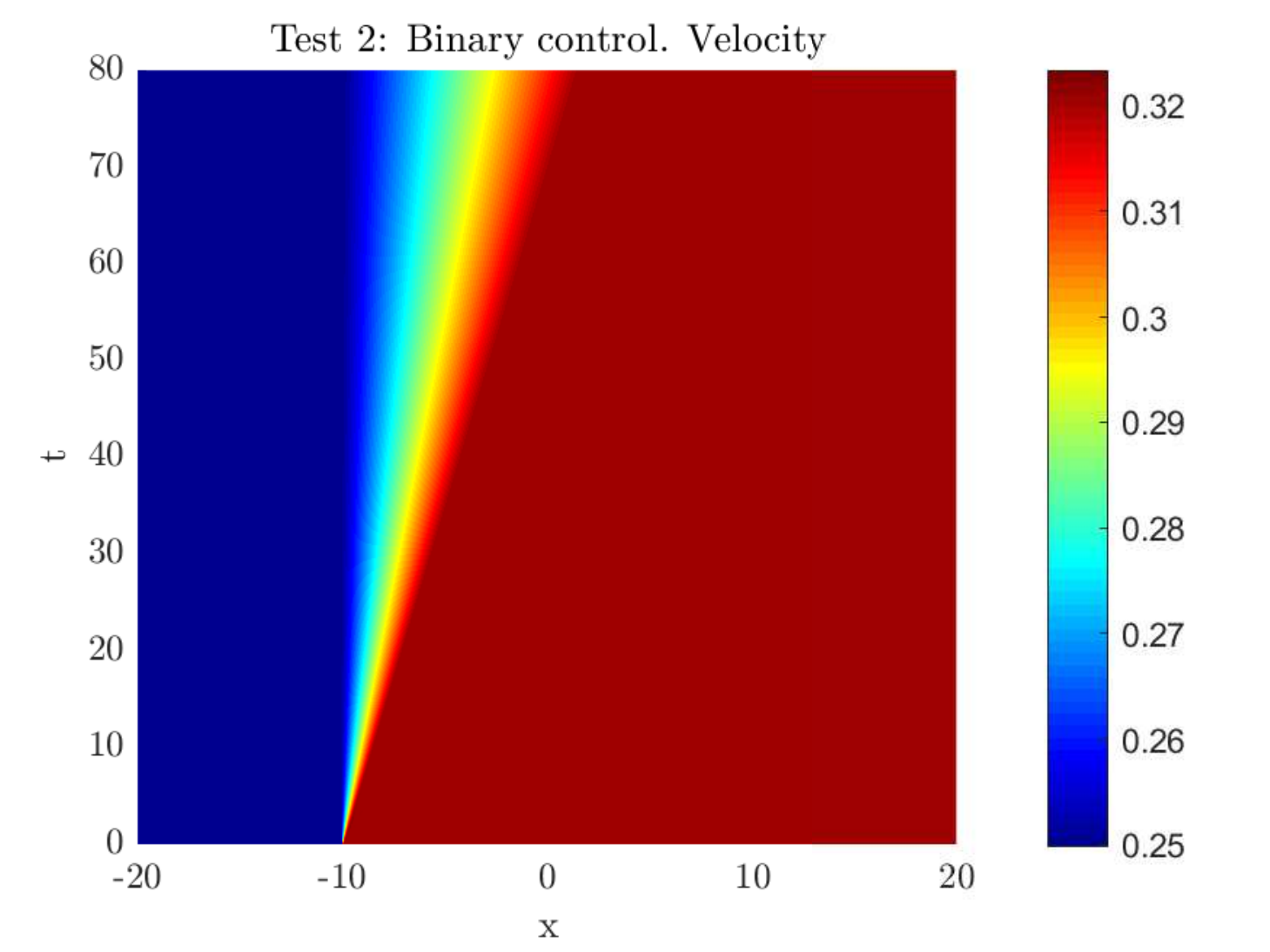}
\includegraphics[width=0.45\textwidth]{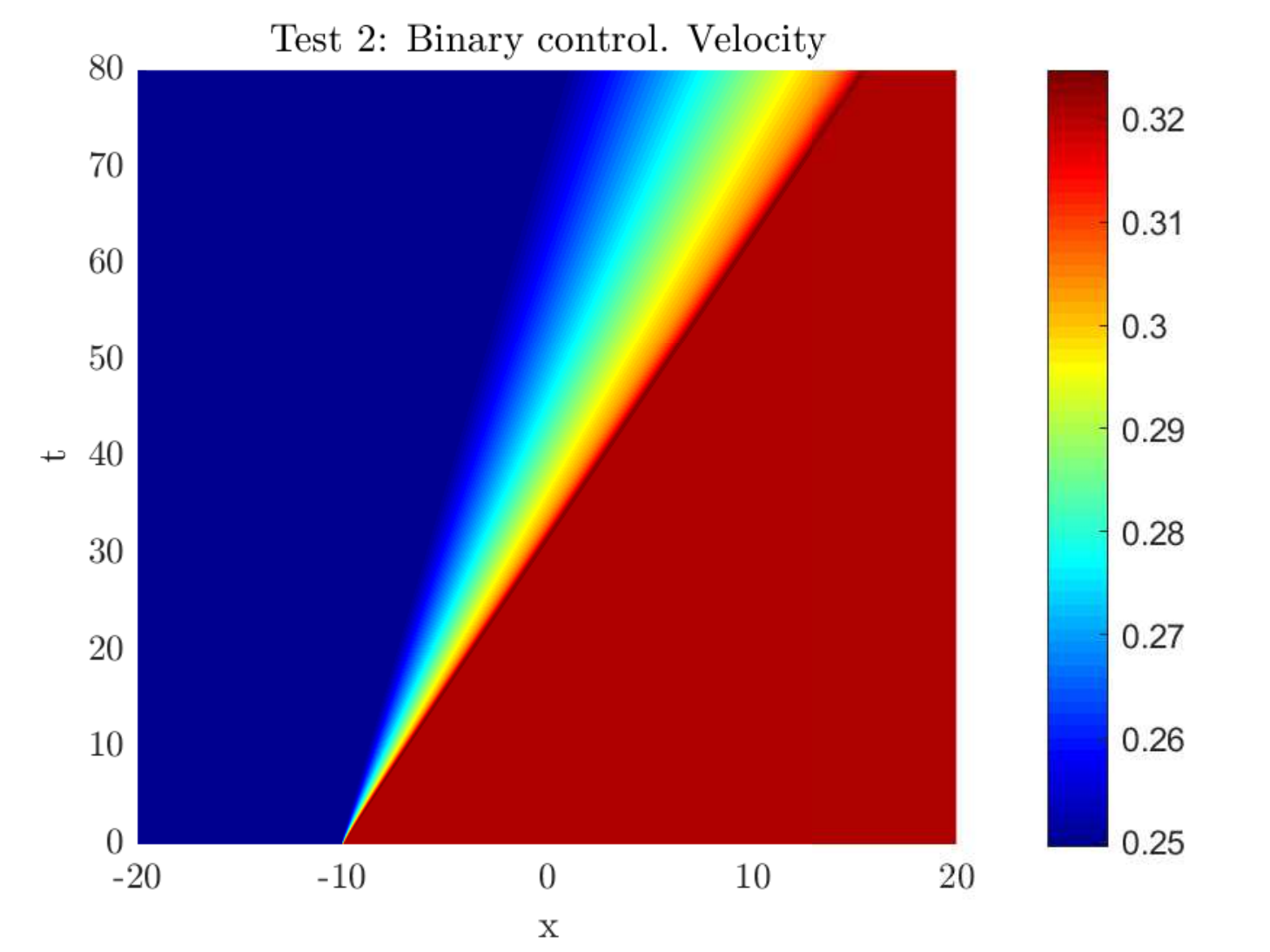}
\caption{\textbf{Test 2B}. Time evolution of the controlled ARZ model with binary control strategy. Top pictures: density, bottom pictures: mean speed. Left pictures show the case $q=0$, right pictures the case $q=1$ with $\lambda(\rho)=\rho$.}
\label{fig:testc_binary3}
\end{figure}

\begin{figure}[t!]
\centering
\includegraphics[width=0.45\textwidth]{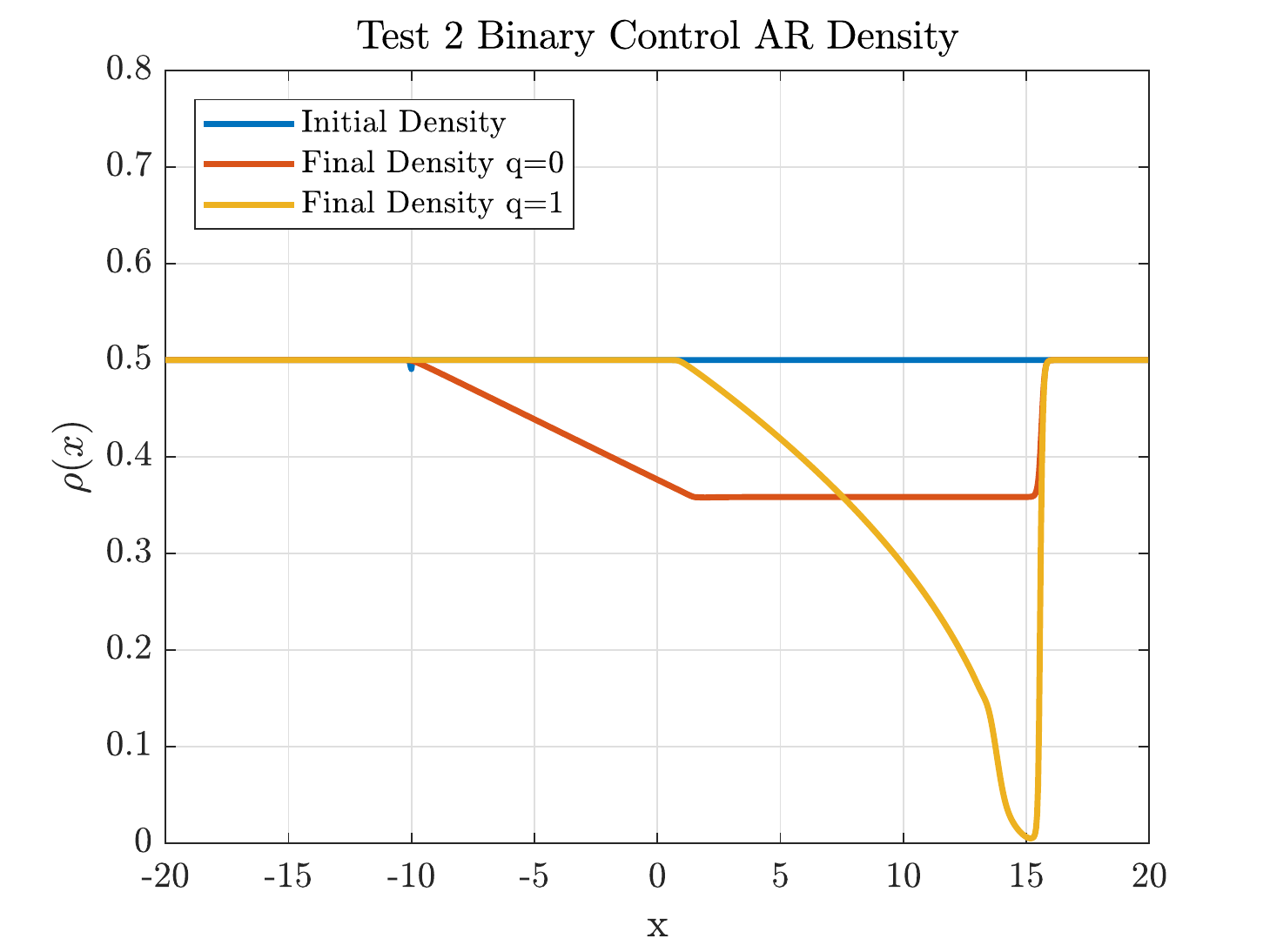}
\includegraphics[width=0.45\textwidth]{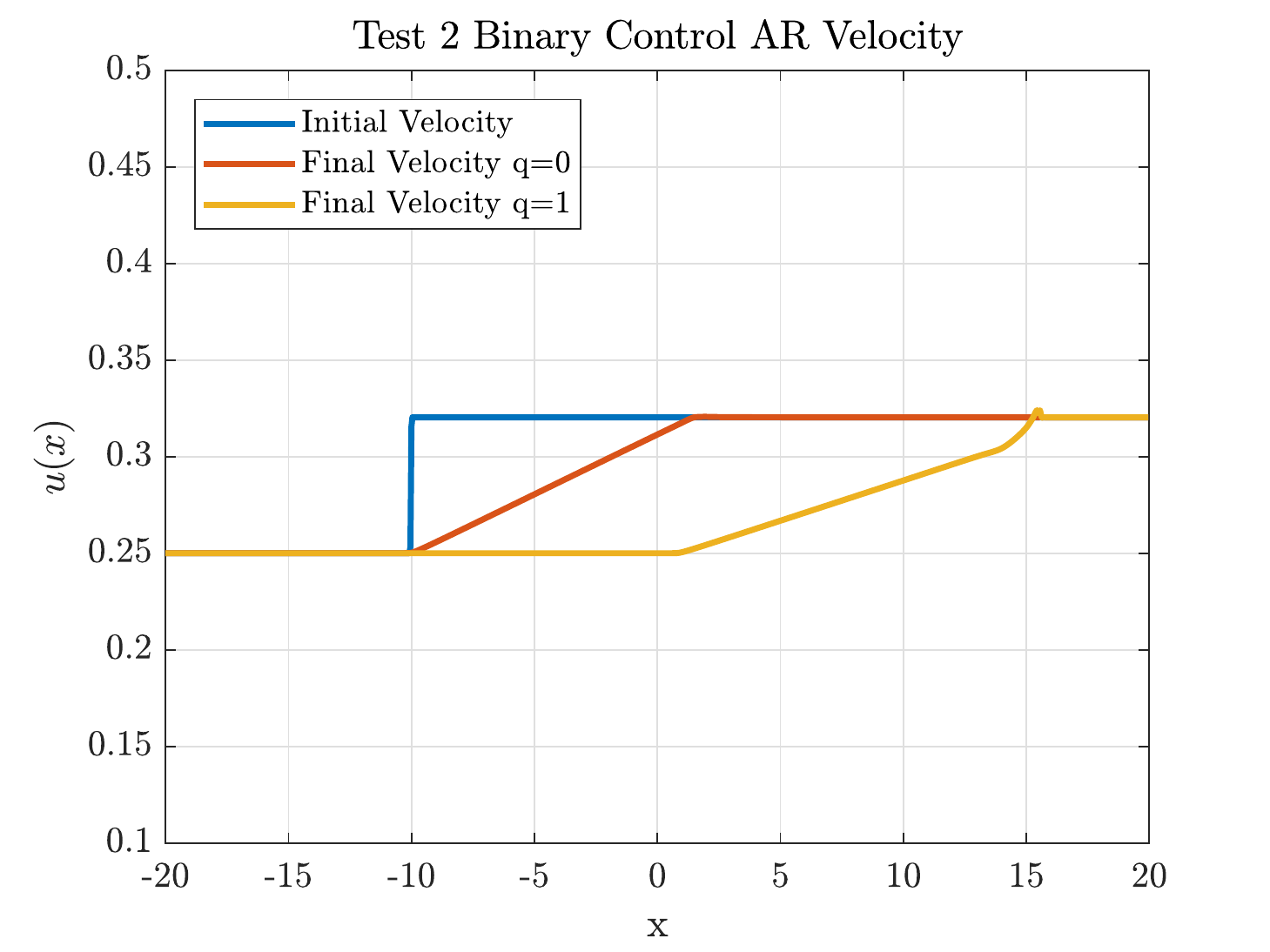}
\caption{\textbf{Test 2B}. Initial and final profile of density and mean speed for the controlled versus the non-controlled ARZ model with binary control strategy. Density (left) and speed (right) with $\lambda(\rho)=\rho$.}
\label{fig:testc_binary4}
\end{figure}

Finally, in Figures~\ref{fig:testc_binary3} and~\ref{fig:testc_binary4} we report the results of a different Riemann problem with initial data
$$	\rho_0(x)=0.5,\ -20\leq x\leq 20,
	\qquad
	u_0(x)=
	\begin{cases}
		0.25 & \text{if } x<-10 \\
		0.32 & \text{if } x\geq -10.
	\end{cases} $$
Figure~\ref{fig:testc_binary3} shows the time evolution of the density and mean speed while Figure~\ref{fig:testc_binary4} shows a comparisons between the controlled and the non-controlled cases at final time. The numerical parameters are the same as the previous test. The final time is now $T=80$. In this case, the solution looks like a rarefaction for both density and mean speed followed by a shock. In the non-controlled case $q=0$ we observe a large variation of the density of vehicles due to the fact that the backward propagation of information goes at a lower pace. Conversely, in the controlled case vehicles adapt faster to the larger mean speed assigned as right state, which produces a smaller rarefaction and a stationary density at the end of the rarefaction wave.

\subsection{Test 3: Controlled ARZ model with desired speed control strategies}
In this last section, we discuss the trend of the kinetic ARZ model when a desired speed control mechanism is activated. This causes a reaction term to appear into the momentum equation, which induces the mean speed $u$ of the vehicles to relax towards the externally prescribed $v_d(\rho)$ at rate $2q\gamma^2/(\nu+\gamma^2)$. The goal is to drive the system towards a desired, in principal optimal, mean speed. We point out that the desired speed $v_d(\rho)$ may be prescribed in many different ways in order to optimise different aggregate aspects of traffic. For instance, one may aim to maximise the net flow through a portion of the road or avoid the onset of congestions. Here we do not consider this further optimisation aspect, which would require one to solve an additional optimal control problem at the macroscopic scale constrained by the hydrodynamic traffic equations. For a contribution in this direction, we refer instead the reader to~\cite{chiarello2020MMS_preprint}.

For the numerical tests of this section we consider now following initial data, mimicking a traffic jam in a portion of the road:
$$	\rho_0(x)=
	\begin{cases}
		0.5 & \text{if } x<0 \\
		0.9 & \text{if } x\geq 0,
	\end{cases}
	\qquad
	u_0(x)=
	\begin{cases}
		1 & \text{if } x<0 \\
		0 & \text{if } x\geq 0.
	\end{cases} $$
Vehicles in $x\geq 0$ are stuck, i.e. their speed is zero. Conversely, vehicles in $x<0$ move with unit mean speed. In the non-controlled case, we expect that vehicles tend initially to accumulate at $x=0$ and that subsequently a queue propagates backwards inducing vehicles in $x<0$ to progressively stop.

We consider three different scenarios:
\begin{enumerate*}[label=\roman*)]
\item in the first one we fix the penetration rate to $q=10^{-3}$, namely we assume a very small number of driver-assist vehicles in the traffic stream;
\item in the second one we slightly increase the penetration rate to $q=7.5\cdot 10^{-2}$;
\item in the third one we fix the penetration rate to $q=1$, i.e. we assume that all vehicles in the traffic stream are equipped with driver-assist technologies.
\end{enumerate*}
In all cases, the desired speed is fixed to
$$ v_d(\rho):=\min\left\{\int_{x-10\Delta{x}}^{x+\Delta{x}}\rho(x,t)\,dx,\,1\right\}, $$
which is conceived in such a way that vehicles trapped in high density portions of the road move faster to reduce the congestion. A maximum dimensionless unit speed is assumed.

\begin{figure}[t!]
\centering
\includegraphics[width=0.45\textwidth]{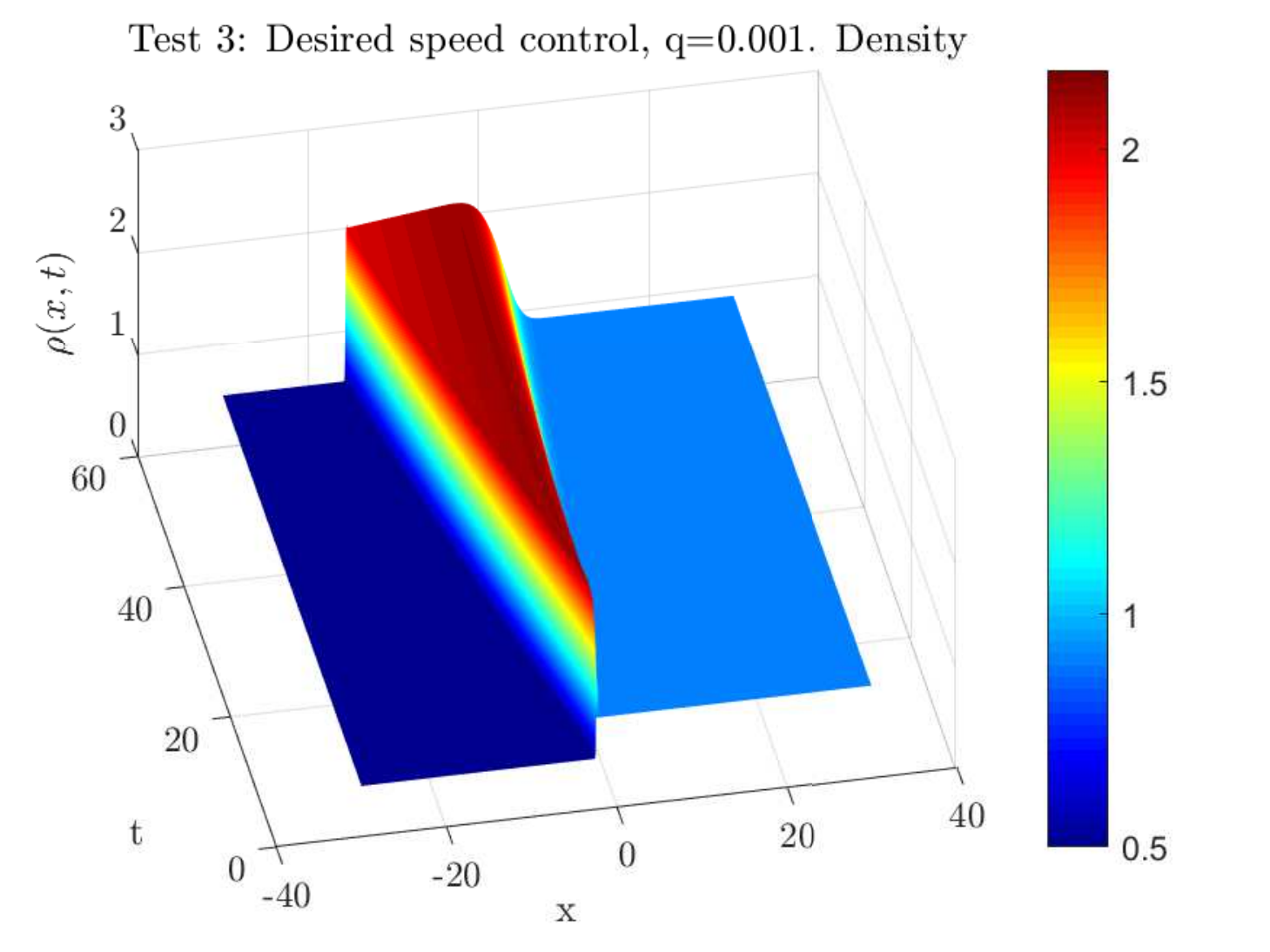}
\includegraphics[width=0.45\textwidth]{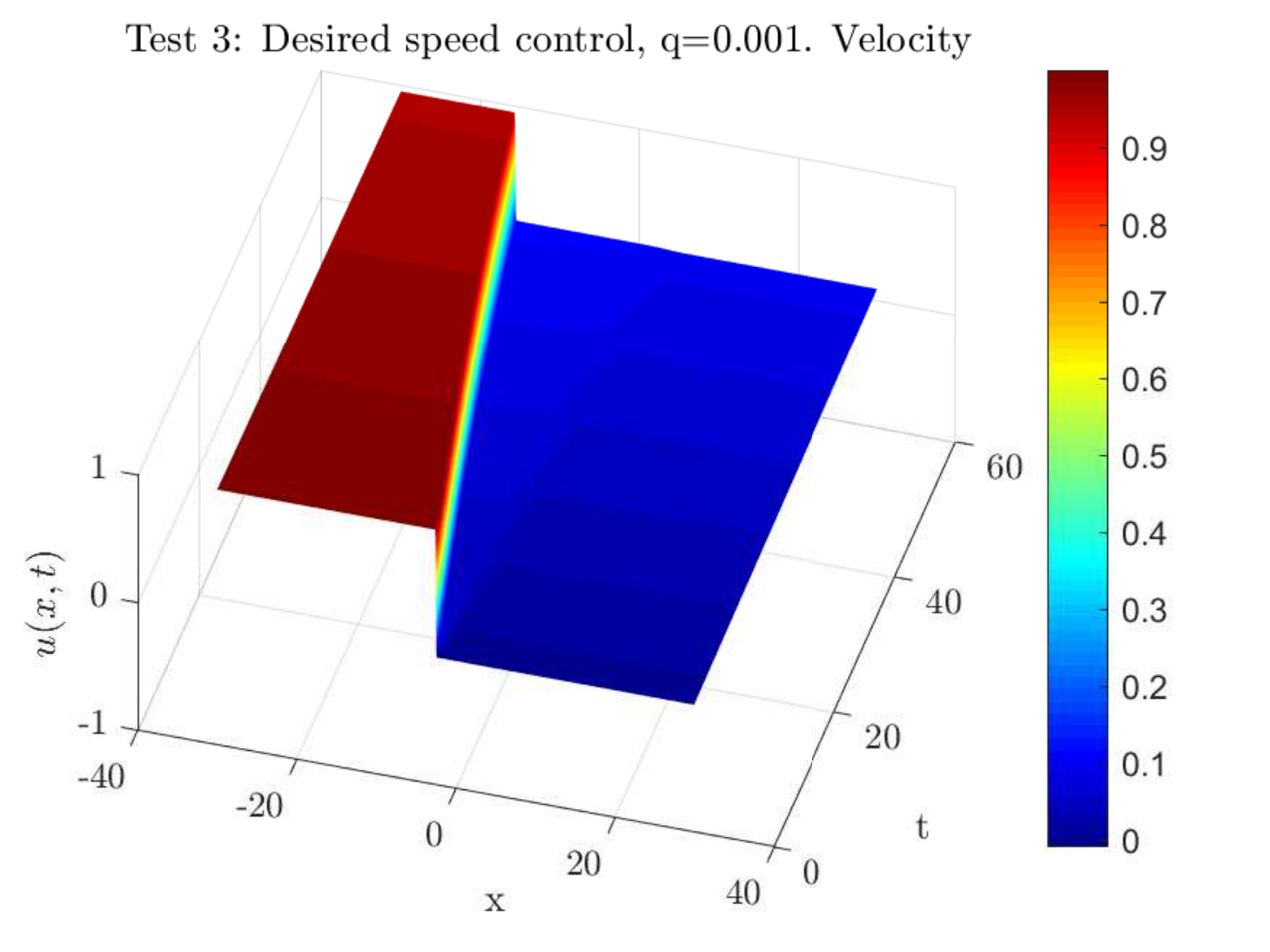} \\
\includegraphics[width=0.45\textwidth]{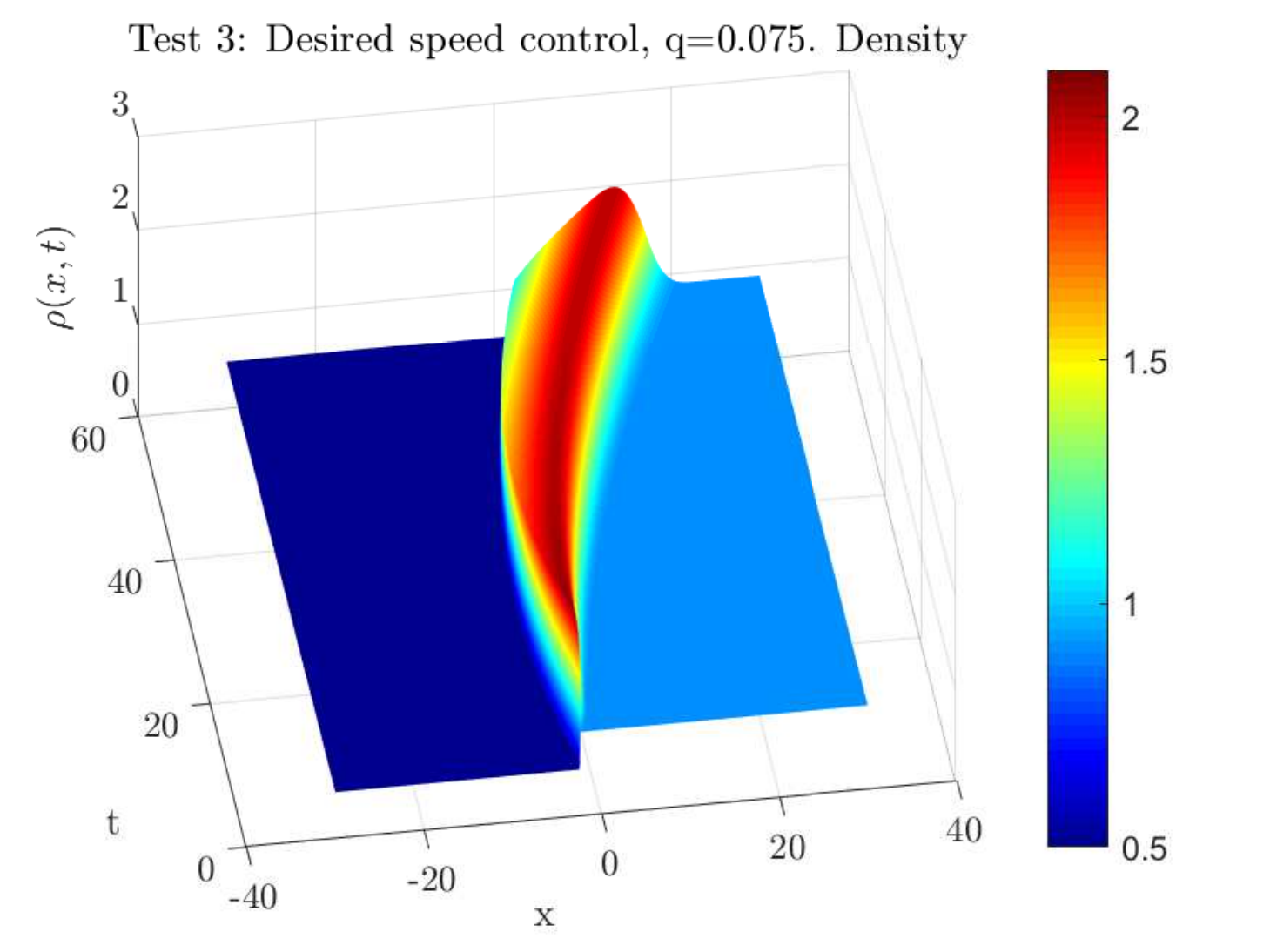}
\includegraphics[width=0.45\textwidth]{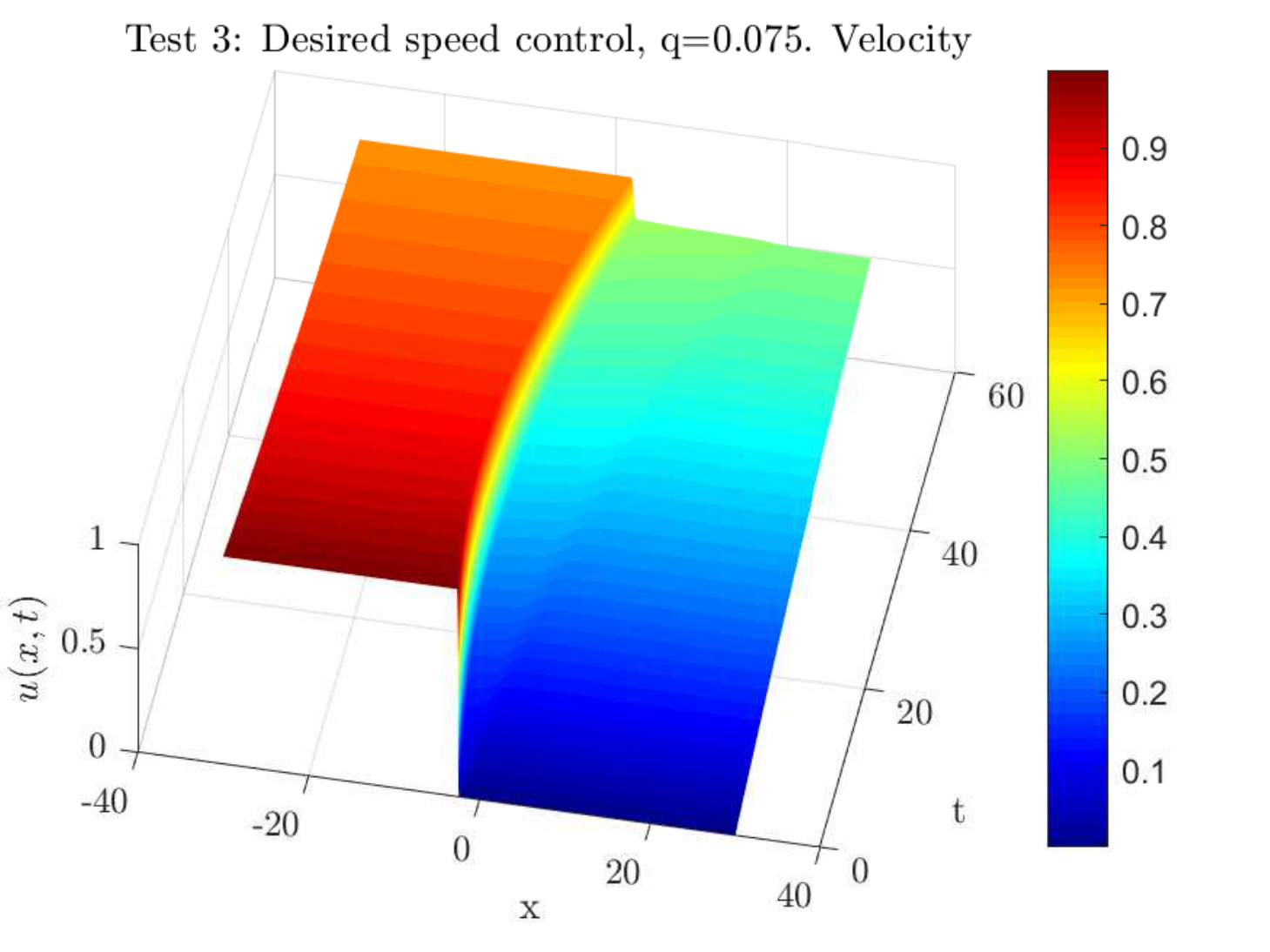} \\
\includegraphics[width=0.45\textwidth]{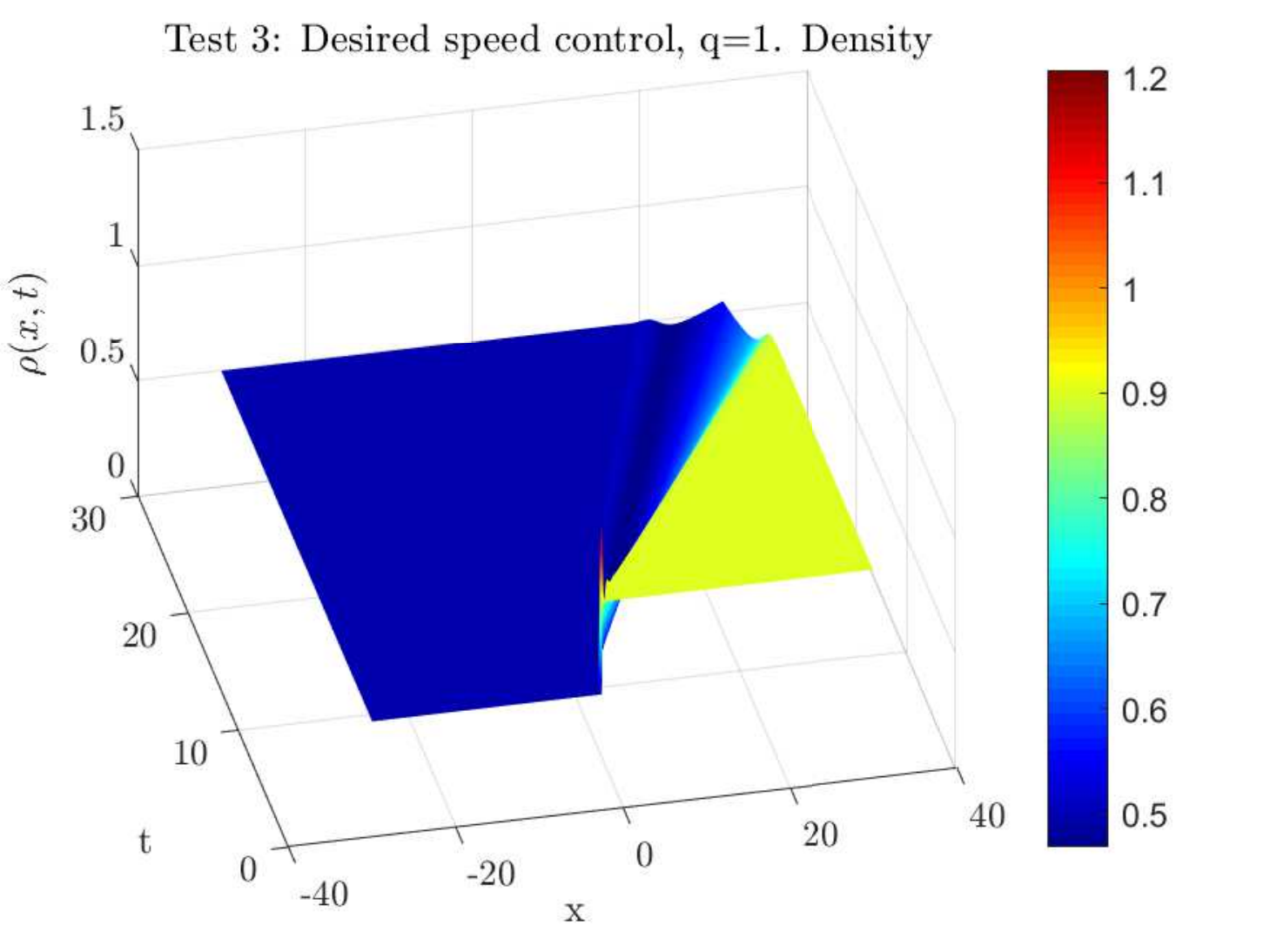}
\includegraphics[width=0.45\textwidth]{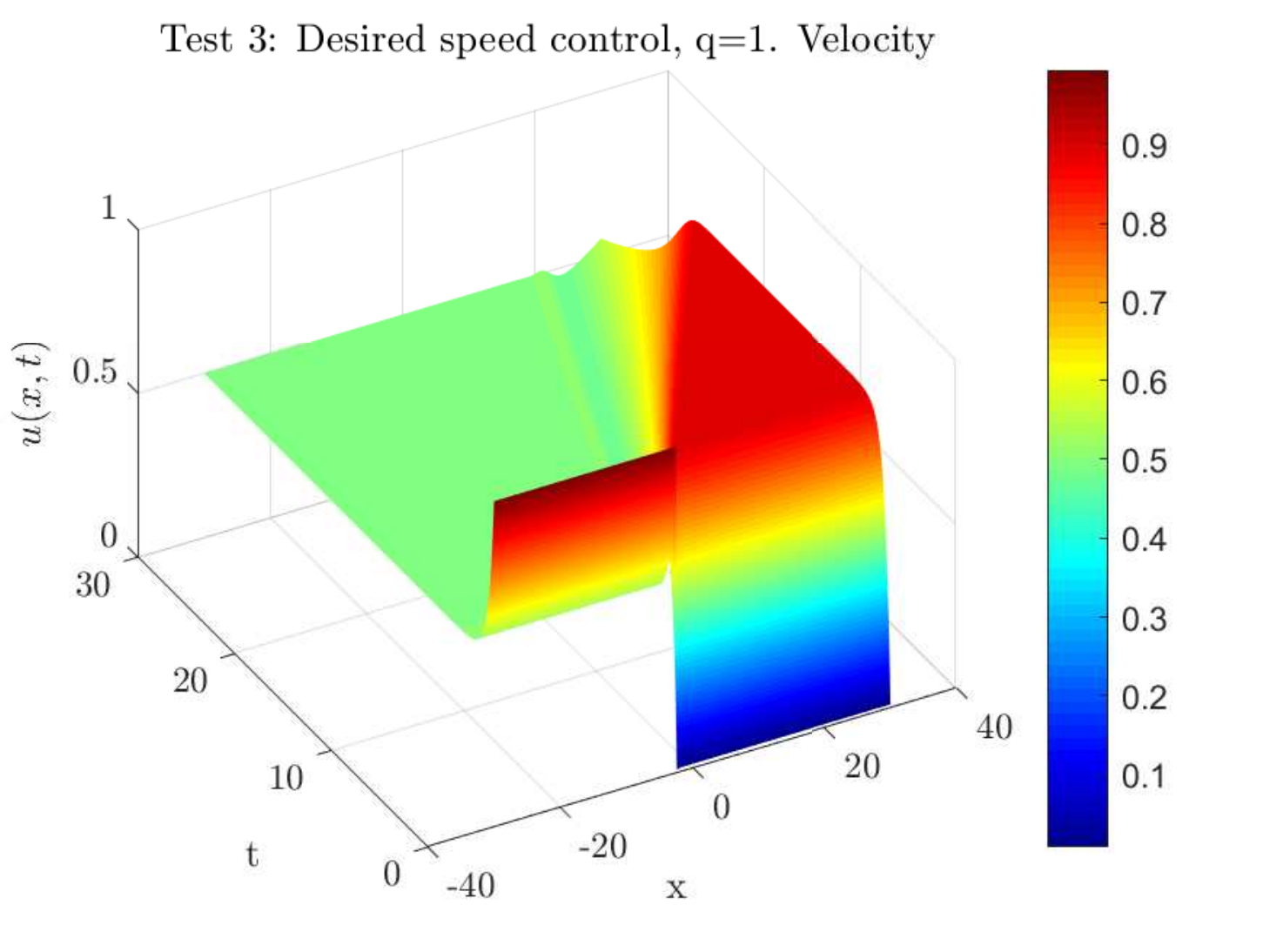} \\
\caption{\textbf{Test 3}. Profiles of density and mean speed in space and time for the kinetic ARZ model with desired speed control. From top to bottom, different penetration rates are considered. Top: $q=10^{-3}$, middle: $q=7.5\cdot 10^{-2}$, bottom: $q=1$. Left pictures: density, right pictures: mean speed with $\lambda(\rho)=\rho$.}
\label{fig:testc_desired1}
\end{figure}

Figure~\ref{fig:testc_desired1} shows the results of the three simulated scenarios in terms of density and mean speed of the vehicles. These results have been computed using $1000$ computational cells over a domain $x\in [-30,\,30]$ by means of the conservative numerical scheme detailed in Section~\ref{sect:test1}. Left pictures represent the density profiles in time and space while right pictures represent the corresponding mean speed profiles. The tests show that a too small percentage of driver-assist vehicles in the traffic stream is unable to prevent the onset and backward propagation of a traffic congestion. In particular, vehicles in $x<0$ slow down in time up to a complete stop. This is indeed the expected outcome also when a completely non-controlled traffic stream is considered. Conversely, in the case $q=1$ we observe that the mean speed of the vehicles in $x\geq 0$ grows more quickly while their density diminishes. In the intermediate case $q=7.5\cdot 10^{-2}$ we notice that, at first, the shock wave in the density starts moving leftwards as for $q=10^{-3}$, which indicates that the traffic jam initially increases in size. Nevertheless, after a while the action of the driver-assist vehicles reverses the dynamics and vehicles start collectively to flow rightwards, thereby preventing the onset and backward propagation of a congestion. However the density peak is only slowly reduced with respect to the case $q=1$.

\section{Conclusions}
\label{sect:conc}
In this paper, we have derived second order hydrodynamic models of ARZ-type describing traffic dynamics in presence of driver-assist controls. Our derivation is formally obtained from the hydrodynamic limit of a combination of local Boltzmann-type and non-local Boltzmann-Enskog-type kinetic descriptions of vehicle interactions. We have considered two different controls motivated by the engineering literature. A first control mimics Adaptive Cruise Control (ACC) devices. A second control is instead inspired by Cooperative Adaptive Cruise Control (CACC) devices, which transmit aggregate information to the vehicles. In the ACC case we have obtained an ARZ-type model with a pressure term modified by the control, whose effect is to make the mean speed of the vehicles more uniform. Instead, in the CACC case we have obtained an ARZ-type model with relaxation towards a prescribed desired speed. Thanks to our approach based on kinetic theory, we have been able to link precisely the key features of these new hydrodynamic models, such as e.g., the form of the modified traffic pressure or of the relaxation term, to structural properties of the interactions among the vehicles and of the feedback vehicle-wise action of the controls. Through targeted numerical experiments we have shown that the obtained hydrodynamic models are able to provide insights into the large-scale dynamics of intelligent traffic streams with no additional analytical and computational costs with respect to the standard fluid dynamics models of non-controlled road traffic. Future research perspectives include the extension of the present models to multilane flows and their use for large-scale traffic optimisation by means of microscopic binary control algorithms.

\section*{Acknowledgments}
This research was partially supported by the Italian Ministry for Education, University and Research (MIUR) through the ``Dipartimenti di Eccellenza'' Programme (2018-2022) -- Department of Mathematical Sciences ``G. L. Lagrange'', Politecnico di Torino (CUP: E11G18000350001) and Department of Mathematics ``F. Casorati'', University of Pavia -- and through the PRIN 2017 project (No. 2017KKJP4X) ``Innovative numerical methods for evolutionary partial differential equations and applications''. This work is also part of the activities of the Starting Grant ``Attracting Excellent Professors'' funded by ``Compagnia di San Paolo'' (Torino) and promoted by Politecnico di Torino.

GD is member of GNCS (Gruppo Nazionale per il Calcolo Scientifico) of INdAM (Istituto Nazionale di Alta Matematica), Italy. AT and MZ are members of GNFM (Gruppo Nazionale per la Fisica Matematica) of INdAM, Italy.

\bibliographystyle{plain}
\bibliography{DgTaZm-ARZ_control}

\end{document}